\documentclass[envcountsame,envcountchap]{svmono}

\usepackage{makeidx}         
\usepackage{graphicx}        
\usepackage{multicol}        
\usepackage[bottom]{footmisc}
\usepackage[pdftex,colorlinks=true,linkcolor=black,citecolor=black,urlcolor=blue]{hyperref}
\usepackage{color} 
\usepackage[numbers]{natbib}
\graphicspath{{./figures/}{./../figures/}}

\newcommand\citer[1]{\citeauthor{#1}\ (\citeyear{#1})\ \cite{#1}}
\newcommand\cites[1]{\citeauthor{#1}'s\ (\citeyear{#1})\ \cite{#1}}
\newcommand\citess[1]{\citeauthor{#1}'\ (\citeyear{#1})\ \cite{#1}}
\newcommand\citeb[1]{[\citeauthor{#1}\ (\citeyear{#1})\ \cite{#1}]}

\usepackage{soul} 
\soulregister\cite7
\soulregister\citer7
\soulregister\cites7
\soulregister\citess7
\soulregister\citeb7
\soulregister\ref7

\makeindex

\begin{document}

\mainmatter
\setcounter{chapter}{1}
\noindent
\textbf{\huge  Filling in the gaps of the tsunamigenic sources in 2018 Palu Bay tsunami}
\vspace*{0.5cm}\\
{\Large Pablo Higuera, Ignacio Sepulveda \& Philip L.-F. Liu}

\section*{Abstract}
The causes of 2018 Palu bay (Indonesia) tsunami are still not entirely clear.
There is still an ongoing debate on whether the main cause of the tsunami waves observed was a significant co-seismic tectonic event which occurred underwater or whether it were the multiple landslides detected along the coast and triggered by the earthquake.
Data from the paper by \citer{liu20} in which the bathymetry of the bay was analysed suggests that landslide-induced waves may have contributed significantly to the tsunami. 
However, the data presented was incomplete and the information regarding the starting time, magnitude of these waves and the coastal landslide progression has significant uncertainties.
In this paper we model each landslide-generated wave with the COMCOT model and track the propagation of the waves to understand their individual contribution at several relevant locations inside the bay where free surface elevation data is available.
We then explore the feasible scenarios (i.e. landslide-generated wave configurations and timings) that produce tsunami waves as close as those observed using an optimization technique based on genetic algorithms.
Numerical simulations of the chosen scenario point out that landslide-generated waves are likely the main contributors to the tsunami, as they can arrive very fast, at the precise timing, to the locations of interest and can trigger the natural resonant modes of the bay, producing long-period waves that were also observed.

\vspace*{3cm}
This is a preprint of the following chapter: Higuera, Sepulveda \& Liu (2021), Filling in the gaps of the tsunamigenic sources in 2018 Palu Bay tsunami, published in Civil engineering for disaster risk reduction, edited by Kolathayar, Pal \& Bharadwaz (2021), Springer.
Reproduced with permission of Springer.
The final authenticated version will be available online at: \url{http:// dx.doi.org/...}

\clearpage

\section{Introduction}
\label{sec:intro}

\begin{figure}
\centering
\includegraphics[width=\textwidth]{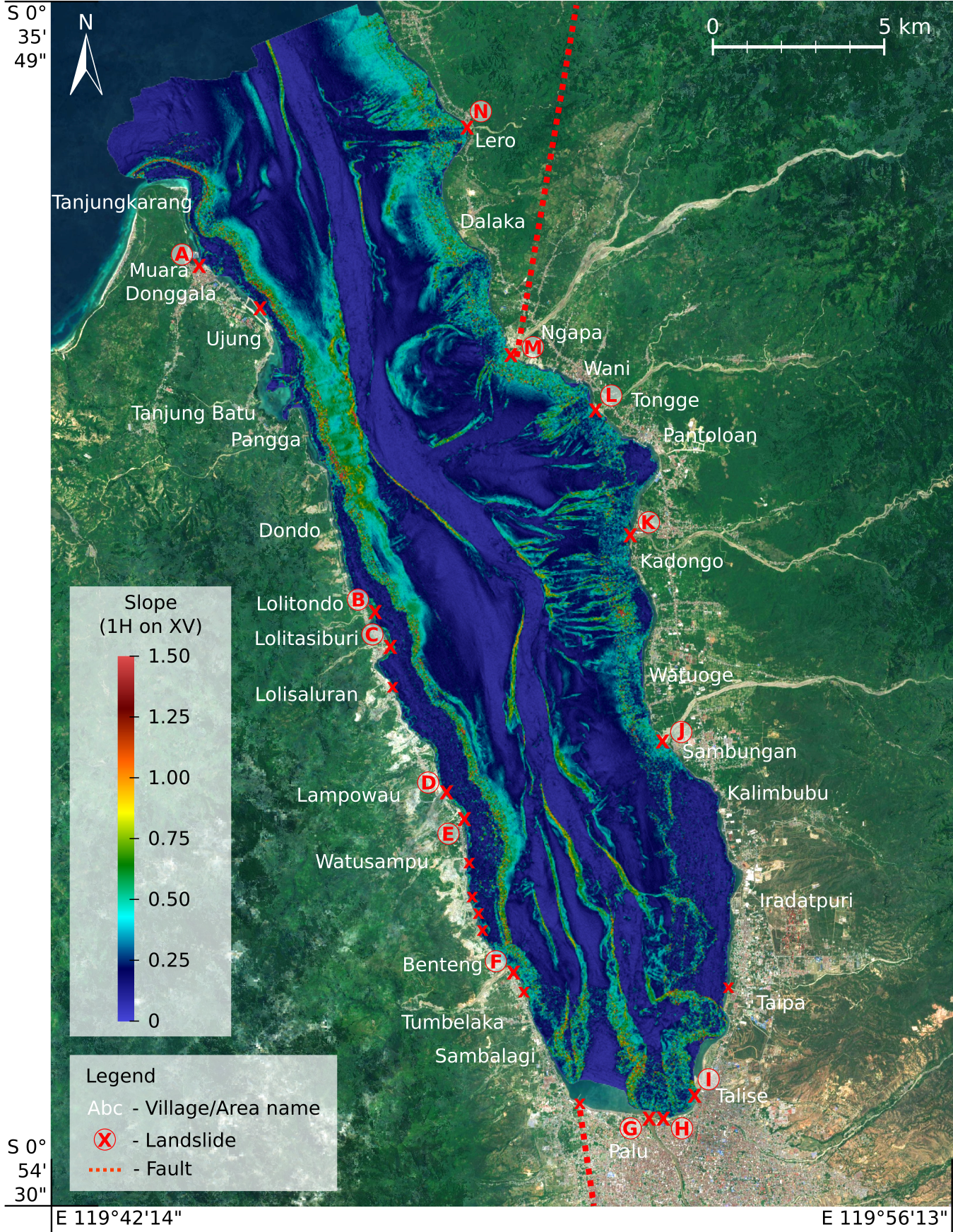}
\caption{Map of Palu bay and slopes of the bathymetry. The coastal landslides are marked with a red cross, and the most relevant ones named with letters from A to N. Figure after \cite{liu20}.}
\label{fig:map}
\end{figure}

On September 30th, 2018, an Mw 7.5 strike-slip earthquake struck towns and villages within Palu Bay, accompanied by landslides and a local tsunami a couple of minutes later.
The death toll of the event has been estimated in approximately 4,340 people.
The infrastructural damage was also significant, as reported by post-tsunami surveys \cite{arikawa18, omira19, paulik19, liu20}.
The map of Palu bay is included for reference in Figure~\ref{fig:map}.
Names of the coastal villages/cities which may be used in this paper are included in white and the mapped Palu-Koro fault line, extracted from \citer{valkaniotis18}, is shown in red line.

While the earthquake magnitude was not a surprise in this tectonically active region with similar events in the past \cite{watkinson17, prasetya01, bellier01}, the peculiar supershear earthquake rupture process \cite{mai19}, the unexpectedly-high local tsunami waves and the possible relationship of both phenomena prompted a long-lasting research effort which remains until today.
Preliminary simulations right after the event \cite{heidarzadeh19, sepulveda18} showed that the tsunami initial condition described by static co-seismic deformation of the bay seafloor cannot explain the size, periods and arrival times observed by eye-witnesses and inferred from post-tsunami surveys.
Two possible sources were hypothesized then as the cause of such a devastating tsunami.
First, the complex geometric and dynamic characteristics of the earthquake rupture caused the large tsunami waves (e.g. \cite{song18, song19, ulrich19}).
Second, several landslides near the bay coastline and triggered by the earthquake contributed to high tsunami waves at different locations.
Hypothetical submarine landslides in the deep end of the bay have also been suggested \cite{heidarzadeh19, pakoksung19}, but they are likely a secondary factor, since there is no evidence of large submarine landslides in the most recent bathymetry surveys \cite{liu20}.
It must be noted that some authors use the term submarine landslide to refer to the coastal landslides, even if these are often only partially submerged; in this work we will only use the term submarine landslides to indicate those which may exist in the deep end of the bay.
In short, the validation of the hypotheses in the early days was extremely challenging due to the scarce amount of local earthquake and tsunami data.

The earthquake information collected for this event includes tele-seismic data, InSAR data and photo imagery in the surroundings of the Palu Bay \cite{bao19, soquet19, sepulveda20}.
While the static modeling of the earthquake co-seismic deformation is straightforward by means of classic dislocation models (e.g. \cite{okada85}), the data is located far from the bay seafloor and a unique fault geometry and earthquake slip distribution cannot be solved \cite{sepulveda20}.
Furthermore, although surface ruptures have been thoroughly studied \cite{jaya19}, the contribution of secondary faults beneath the bay or the possibility of a complex dynamic contribution of the seafloor displacement to the tsunami is still unclear, as the deepwater portion of the fault remains unmapped.
Information about the tsunami was firstly provided by a tide gauge located at the port of Pantoloan, inside the bay, with a sampling interval of 1 minute.
As shown by eye-witness videos, the first and most destructive tsunami waves were short and, therefore, the available gauge data could only provide partial information of the tsunami evolution.
\citer{carvajal19} used the video footage from social media to reconstruct tsunami time histories at different locations within Palu Bay.
The footage evidenced the issues of the tide gauge capturing the first tsunami waves at Pantoloan and provided further information to constrain earthquake and tsunami models.
This study and several others (e.g. \cite{arikawa18, sassa19, omira19}) also evidenced signs of localized subsidence at the coast, presumably along with submarine landslides.
All these datasets were supported by post-tsunami surveys which provided key information about the inundation and coastal subsidence \cite{omira19, fritz18}.
The partial earthquake and landslide information from these datasets were fully exploited in new tsunami simulations using a variety of models and approaches \cite{schambach21}.

Until today, the studies have arrived to disparate conclusions, the landslides being the primary source \cite{takagi19, liu20} or a combination of both landslide and co-seismic deformation (e.g. \cite{aranguiz20, schambach21}).
Early works, e.g. \cite{ulrich19}, also hypothesised with the co-seismic deformation being the main source of the tsunami, however, the authors are not aware of any new references using all the new data available presently to support this conclusion.
In that sense, the three options seem plausible from a theoretical point of view, since all these mechanisms are known to be capable of producing tsunamis and have been observed before, e.g., in the 1998 Papua New Guinea \cite{synolakis02, harbitz06} or 2010 Haiti tsunamis \cite{fritz13}.
The key in this analysis should be, in the authors' opinion, defining precisely the meaning of ``main'', since some degree of contribution of both factors is highly likely.
However, we also acknowledge that given the major uncertainties in the existing data, ``minor'' contributions may be significantly diluted or amplified because of them.

While multiple landslides were identified along the coastline of Palu from post-tsunami surveys and photo imagery, their actual size, tsunamigenic potential and triggering time and evolution are not well constrained.
Optimization methods have been presented to characterize the landslide sources from the available tsunami data ranging from trial-and-error to inversion methods.
\citer{sepulveda20} employed a linear least square inversion to solve for the tsunami initial elevation at photo-identified landslide locations.
The inversion results suggested that the triggering time and evolution of the landslides play a critical role to fit the observations from video footage and post-tsunami survey inferred runups.
The delayed triggering time and finite duration of landslides were evident from the video footage analysis \cite{carvajal19, sepulveda20}.
A set of co-seismic deformation models inferred from InSAR data were also combined with the inverted landslides tsunami initial conditions.
The simulations showed a minor contribution of the earthquake co-seismic deformation as compared to the landslide sources.
A step-forward to accurately identifying the source of the Palu tsunami is the inclusion of more geological data and the introduction of more sophisticated landslide tsunami models.
\citer{liu20} conducted a thorough bathymetry survey within Palu Bay, including shallow areas where landslides possibly occurred.
The study characterized the possible dimensions of the landslides which constitute key information to further constrain landslide tsunami simulations.
The high-resolution of the bathymetry also provided complementary information of the fault system beneath Palu Bay \cite{natawidjaja20}.

In this study, we aim to use the new landslide data inferred by \citer{liu20} and solve for the landslide conditions and timing which better fit tsunami observation by solving an optimization problem based on Genetic Algorithms (GA) \cite{mitchell98} and numerical simulations performed with the nonlinear shallow water solver COMCOT \cite{wang09}.
In this chapter, we present preliminary results.
Section~\ref{sec:methodology} introduces the methodology to estimate the missing characteristics of the landslides and the waves that they produce.
The following section discusses the most relevant findings from the new simulations.
Finally, Section~\ref{sec:conclusions} draws the final conclusions and describes the future worklines.

\section{Methodology}
\label{sec:methodology}

This work aims to complement and extend the previous analysis performed in \citer{liu20}, most of which was condensed in Table~2 of that work, and partially reproduced in this work in Table~\ref{tab:landslideCharacteristics}.

In \citer{liu20} a detailed bathymetry of Palu bay, acquired after the event, was compared with existing, less detailed, bathymetries from 2014-2017.
Pre- and post-earthquake satellite images were also compared.
The analysis allowed to detect all coastal landslides and to characterize their relevant magnitudes, such as the emerged and submerged dimensions or the volume.
Based on that data, the landslide-generated wave (LGW) characteristics were calculated using the semi-empirical formulation by \citer{lo17}, and were finally simulated using the nonlinear shallow water model COMCOT \citer{wang09}.

The data presented in \citer{liu20} had significant sources of uncertainty (see SM Text 1 in \cite{liu20} for the full list) and some data items were missing because they could not be inferred from the data available.
Table~\ref{tab:landslideCharacteristics} depicts this situation.
The data in black has been extracted from \citer{liu20}.
$L_{ls}$ and $L_{cs}$, and $E_{ls}$ and $E_{cs}$ are the emerged and total (emerged + submerged) dimensions of the landslide, respectively, in the longshore (ls) and crosshore (cs) directions, in metres.
$A$ is the surface area lost, in hectares (10,000 m$^2$) and $V$ is the total volume of the landslide, in millions of m$^3$.
$H$ is the wave height produced by the landslide and $\lambda$ is the wavelength, both in metres.
Finally, $t_s$ is the starting time of the LGW after the earthquake start, in seconds.
This time does not indicate when the coastline collapses, but instead, when the LGW has fully developed and become independent of the landslide, as defined in \citer{loPhD}.

\begin{table}
\centering
\caption{Landslide characteristics, partially reproduced from \citer{liu20}. Data in blue has been estimated in the present work. Data with an asterisk indicates a significant level of uncertainty.}
\label{tab:landslideCharacteristics}
\begin{tabular}{ccccccccccccccc}
\hline\noalign{\smallskip}
 & \textbf{A} & \textbf{B} & \textbf{C} & \textbf{D} & \textbf{E} & \textbf{F} & \textbf{G} & \textbf{H} & \textbf{I} & \textbf{J} & \textbf{K} & \textbf{L} & \textbf{M} & \textbf{N} \\
\noalign{\smallskip}\hline\noalign{\smallskip}
$L_{ls}$ & 255 & 295 & 430 & 1100 & 475 & 705 & 725 & 505 & 250 & 440 & 430 & 460 & 700 & 615 \\
$L_{cs}$ & 72 & 125 & 40 & 135 & 75 & 115 & 90 & 85 & 90 & 140 & 70 & 160 & 125 & 70 \\
$A$ & 1.17 & 1.82 & 0.89 & 6.40 & 1.39 & 3.23 & 2.56 & 2.83 & 1.59 & 3.21 & 1.30 & 6.30 & 4.94 & 1.62 \\
$E_{ls}$ & 260 & 380 & 440 & 1220 & 335$^*$ & 755 & \textcolor{blue}{670} & \textcolor{blue}{435} & 310 & \textcolor{blue}{370} & \textcolor{blue}{360} & 515 & 350 & 390$^*$ \\
$E_{cs}$ & 350 & 340 & 405 & 410 & 175$^*$ & 340 & \textcolor{blue}{280} & \textcolor{blue}{470} & \textcolor{blue}{390} & \textcolor{blue}{620} & \textcolor{blue}{275} & 830 & 800 & 295$^*$ \\
$V$ & 0.41 & 1.44 & 2.26 & 3.07 & 0.37$^*$ & 2.22 & \textcolor{blue}{1.79} & \textcolor{blue}{2.03} & 2.54$^*$ & 0.67$^*$ & 1.11$^*$ & 6.66 & 3.44 & 0.36$^*$ \\
$H$ & 3.9$^*$  & 5.8 & 5.2 & 10.2 & 2.5 & 13.6 & \textcolor{blue}{6.2} & \textcolor{blue}{6.8}  & \textcolor{blue}{8.1} & \textcolor{blue}{1.8} & \textcolor{blue}{2.9} & 11.1 & 4.2 & \textcolor{blue}{1.4} \\
$\lambda$ & 118 & 658 & 1618 & 644 & 252 & 511 & \textcolor{blue}{788} & \textcolor{blue}{808} & \textcolor{blue}{834} & \textcolor{blue}{1054} & \textcolor{blue}{1068} & 1382 & 1997 & \textcolor{blue}{736} \\
\noalign{\smallskip}\hline
\end{tabular}
\end{table}

The objective of this work is to fill the gaps and complete the missing data Table~\ref{tab:landslideCharacteristics}, shown  in blue font.
In order to do so we analyse the existing landslide data and develop ad-hoc scaling laws for the landslides in Palu bay to estimate the missing values (Section~\ref{sec:dataAnalysis}).
With the existing and new LGW characteristics we simulate the propagation of each tsunami wave individually and analyse the effects that they cause all over Palu Bay and the influence at several locations in which free surface elevation data has been estimated from different sources in \citer{carvajal19} (Section~\ref{sec:propagation}).
Finally, empirical adjustments to the wave height, wave length and the initiation time of the landslide-generated tsunamis are sought using an optimization technique based on GA to identify if it is possible to explain better the free surface elevation data measured/estimated at the key locations (Section~\ref{sec:adjustments}).

\subsection{Analysis of the existing landslide data}
\label{sec:dataAnalysis}

The first step of the methodology is to explore the existing data (in black font in Table~\ref{tab:landslideCharacteristics}) and seek simple nonlinear regression models to develop ad-hoc scaling laws for the coastal landslides.
We only include those landslides for which the data that we use at that stage is complete to forecast the missing magnitudes of the rest.
Moreover, only the translational landslides are accounted for, which is why landslide A is not included in the present analysis.

The procedure that we follow in this work is simple and only requires simple techniques that are completely established in data science.
Another option would have been to perform numerical simulations of the landslides, to simulate their interaction with water and measure the wave conditions directly.
Numerical modelling of LGW has been developed quite extensively, see \citer{romamo20} for a comprehensive review.
However, one of the latest examples in the literature, \citer{ghaitanellis21}, highlights that the computational cost is usually very high and simulations require knowledge of the soil properties to accurately model the landslide mechanics.
Therefore, we believe that the simple approach applied in this work is aligned with the level of uncertainty of the present data and is adequate to provide meaningful insights.

\begin{figure}
\centering
\includegraphics[width=0.49\textwidth]{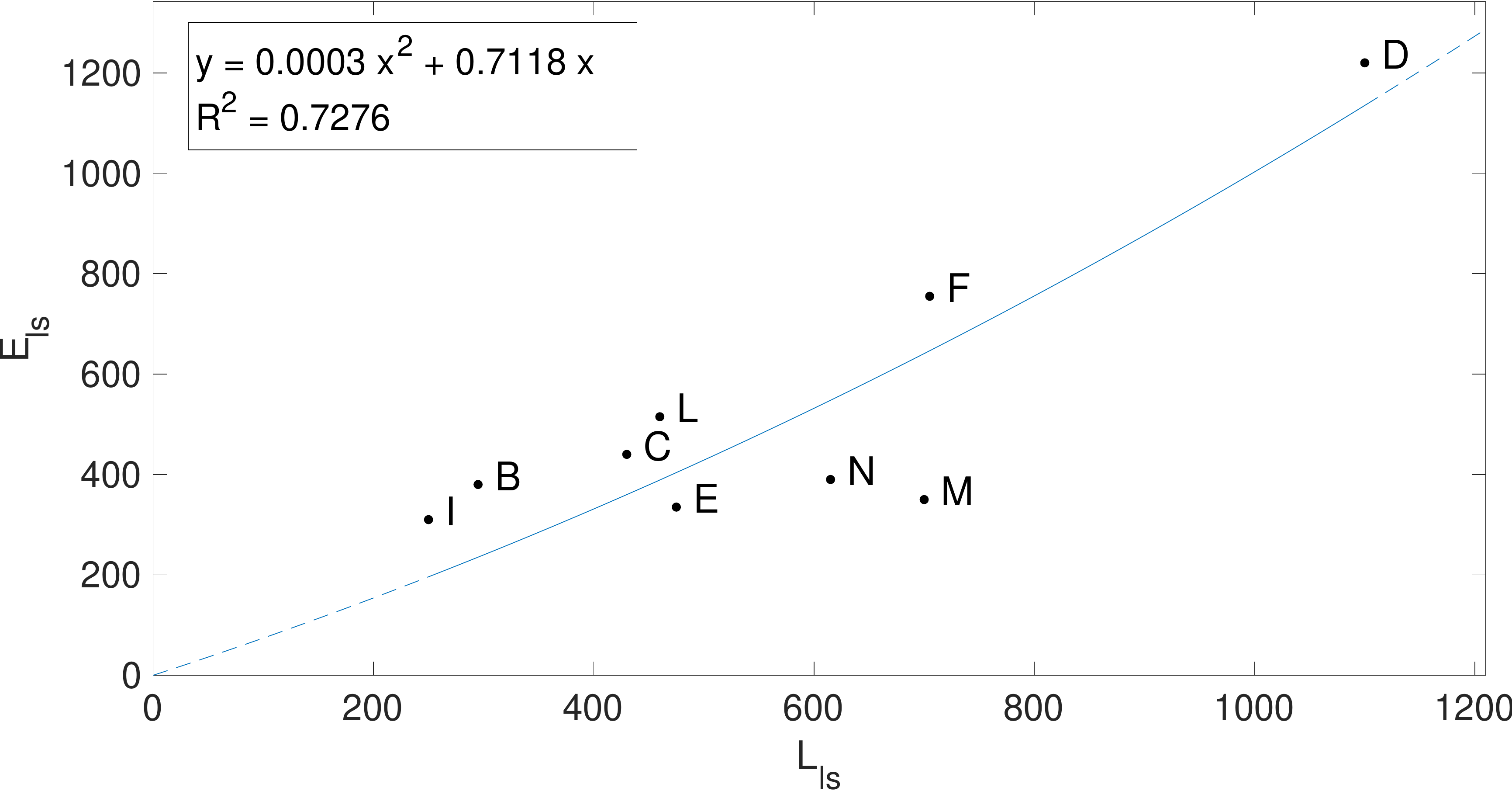}
\includegraphics[width=0.49\textwidth]{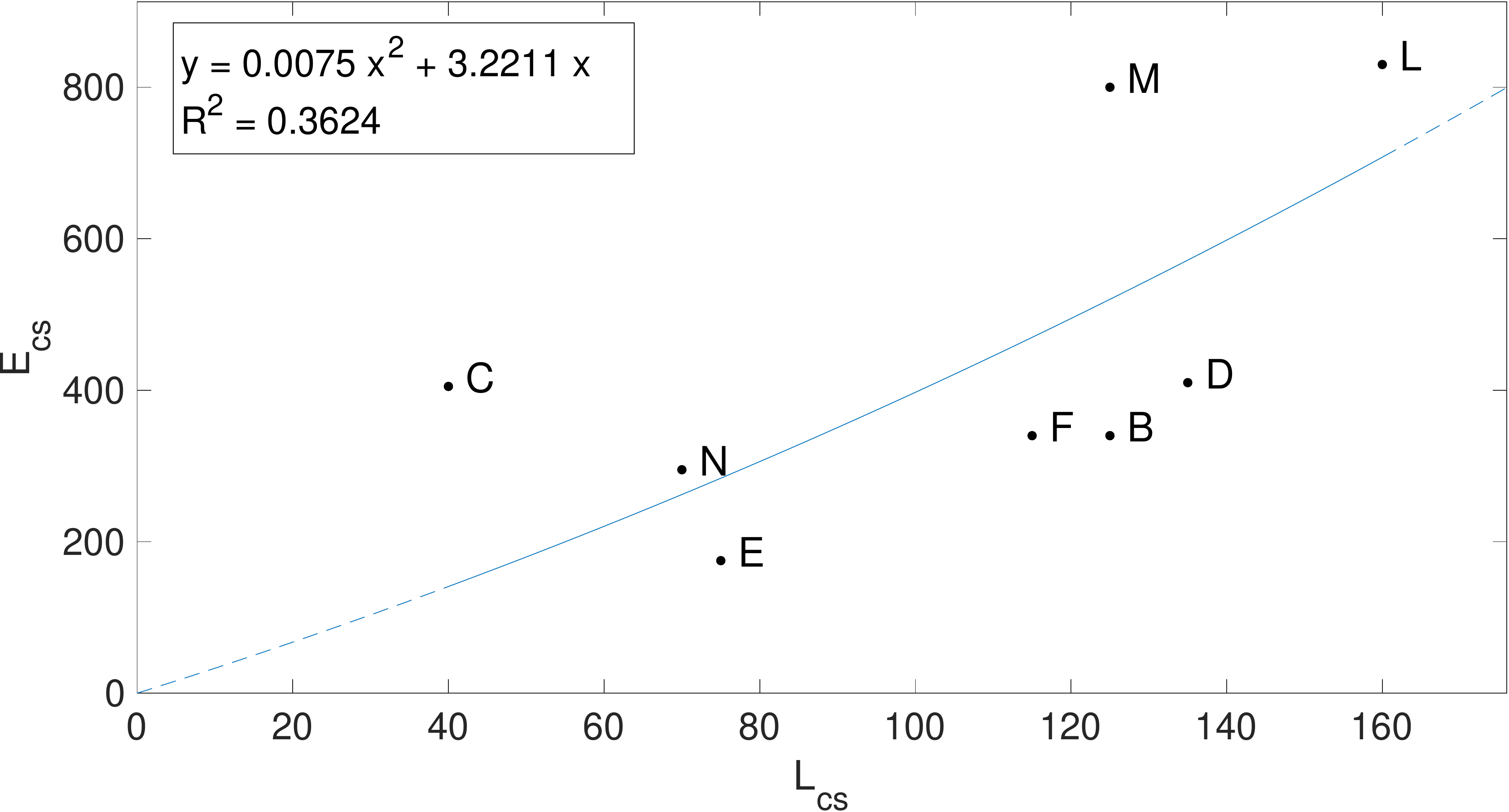}\\
\includegraphics[width=0.49\textwidth]{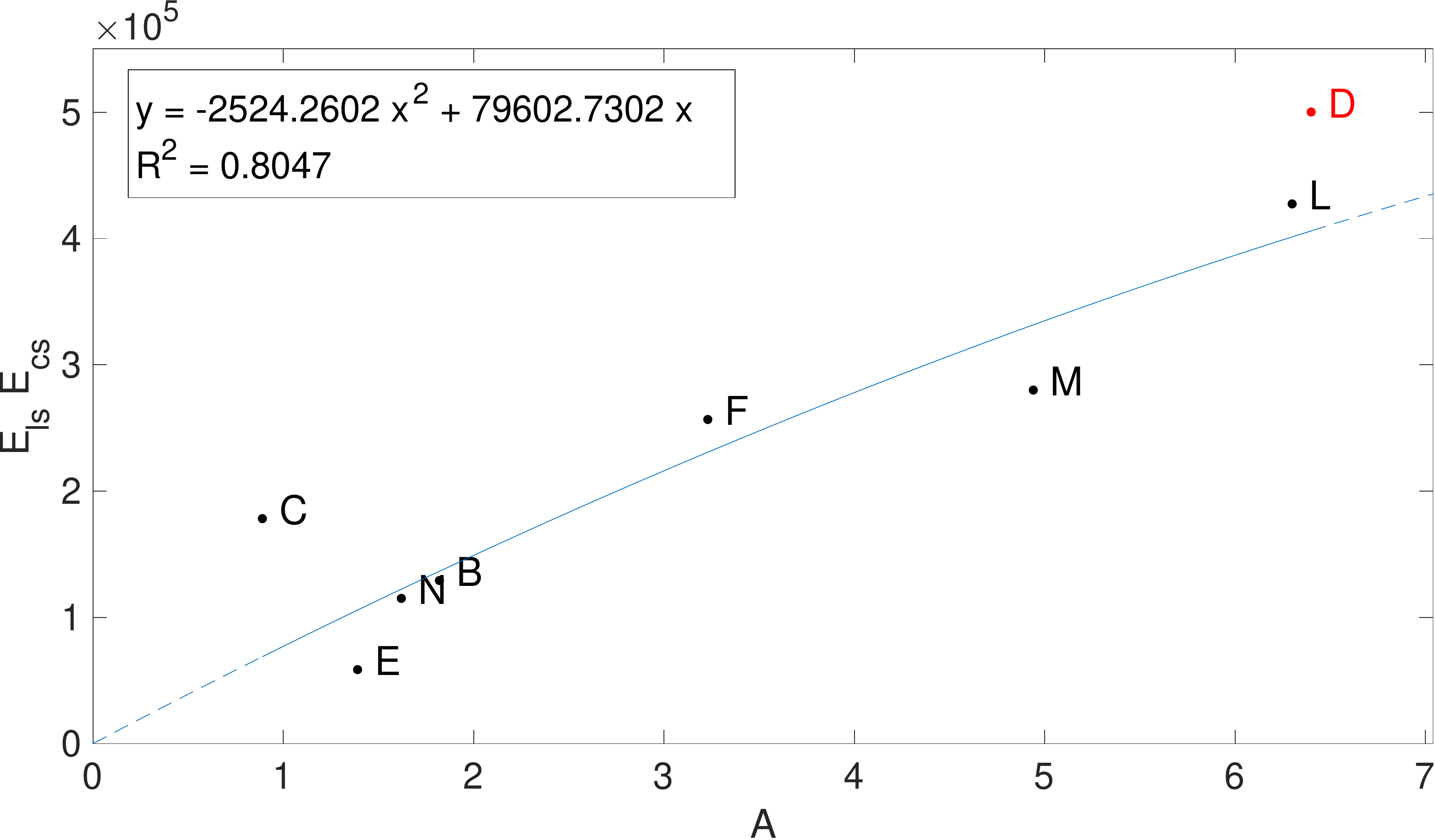}
\caption{Correlation between the emerged and total landslide dimensions (top panels). Correlation between the emerged area lost and the product of total landslide dimensions (bottom panel). Outliers in red.}
\label{fig:correlation1}
\end{figure}

The analysis of the data begins seeking correlations between the emerged dimensions of the landslides ($L_{ls}$ and $L_{cs}$), which are known for all the landslides, and the total dimensions ($E_{ls}$ and $E_{cs}$), some of which are known only partially.
As can be seen in the top left panel in Figure~\ref{fig:correlation1}, there is a fair correlation between $E_{ls}$ and $L_{ls}$ for a parabolic model in terms of the coefficient of determination ($R^2 = 0.7276$).
The model has been chosen to be parabolic because it allows fitting slightly more complex data than a linear model, although in this case the resulting curvature is very small.
Moreover, the parabolic model has been constrained to pass through (0,0), since when $L_{ls}$ approaches to zero (meaning that the emerged landslide vanishes), so will $E_{ls}$.
This constraint is also applied when linking the rest of the variables in this section.
The equation in the top left corner of the panel has been applied to estimate $E_{ls}$ in Table~\ref{tab:landslideCharacteristics} (in blue).

\begin{figure}
\centering
\includegraphics[width=0.49\textwidth]{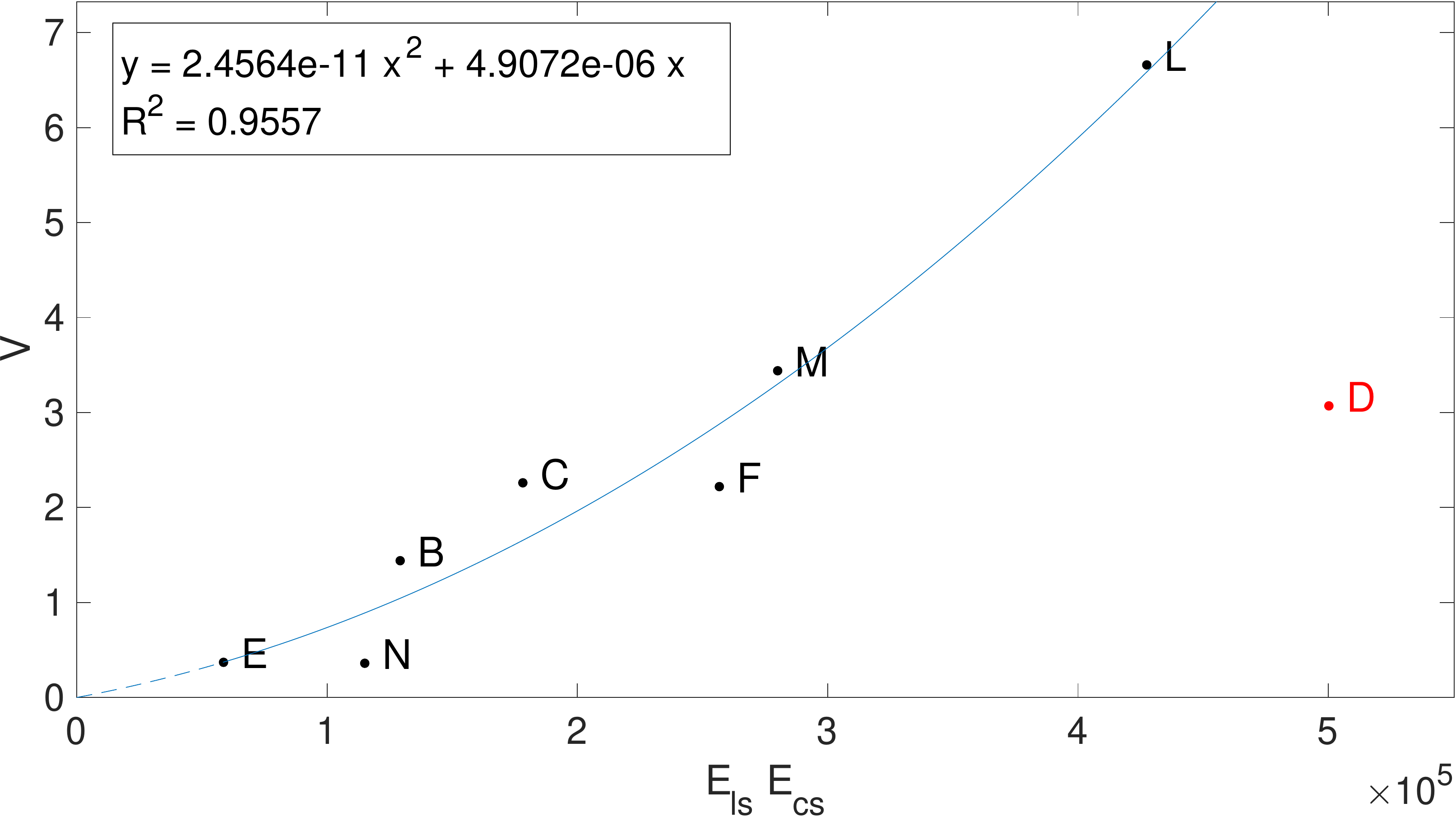}
\caption{Correlation between the product of total landslide dimensions and the volume of the landslide. Outliers in red.}
\label{fig:correlation2}
\end{figure}

The correlation between $E_{cs}$ and $L_{cs}$ (top right panel in Figure~\ref{fig:correlation1}) is quite low ($R^2 = 0.3624$), probably due to the diversity in slopes and local geological aspects; therefore, there is a need to find an alternative to estimate the missing values for $E_{cs}$, without leaving the physics of problem out of sight.
The solution was to try to correlate the land area lost ($A$) with the product of $E_{ls}$ and $E_{cs}$, in the bottom panel.
In this case landslide D is considered an outlier in terms of the product of the total landslide dimensions, as later it will be shown and discussed in Figure~\ref{fig:correlation2}, and as so is represented in red.
Disregarding D, the correlation is very good ($R^2 = 0.8047$) and the equation of the parabola at the top left box of the panel has been used to estimate $E_{cs}$ in Table~\ref{tab:landslideCharacteristics}.

Once the total dimensions of the landslides have been obtained, the next step was to estimate the missing landslide volumes ($V$).
Several combinations of variables were tested, but in the end the best correlation ($R^2 = 0.9557$) was found with the product $E_{ls} \cdot E_{cs}$, as shown in Figure~\ref{fig:correlation2}.
Observing this figure it is obvious why landslide D has been deemed an outlier in terms of $E_{ls} \cdot E_{cs}$, as it is too far away from its closest volumetric neighbours (C, F, M).
The missing landslide volumes ($V$) are obtained using the parabolic expression, and included in blue font in Table~\ref{tab:landslideCharacteristics}.

\begin{figure}
\centering
\includegraphics[width=0.49\textwidth]{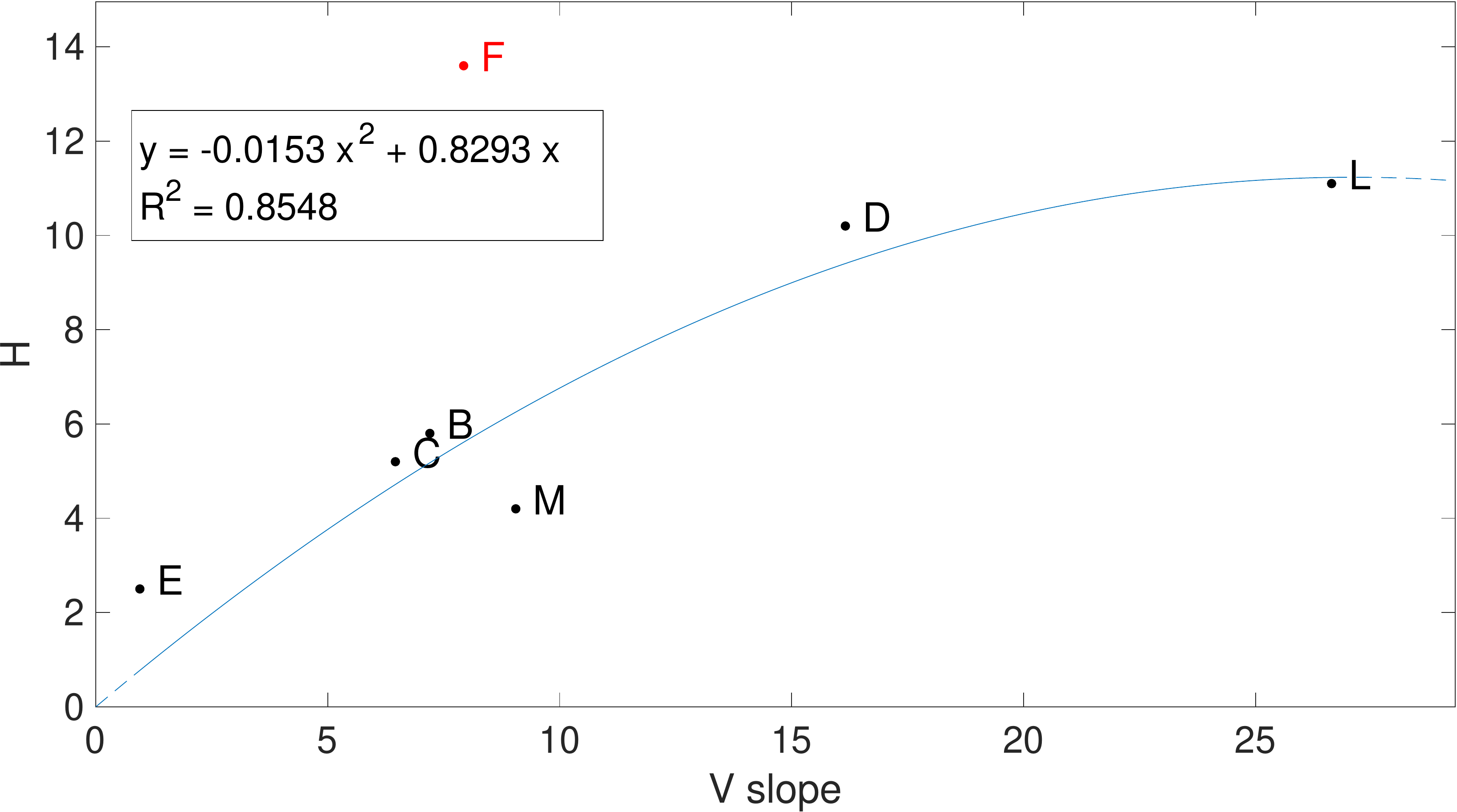}
\includegraphics[width=0.49\textwidth]{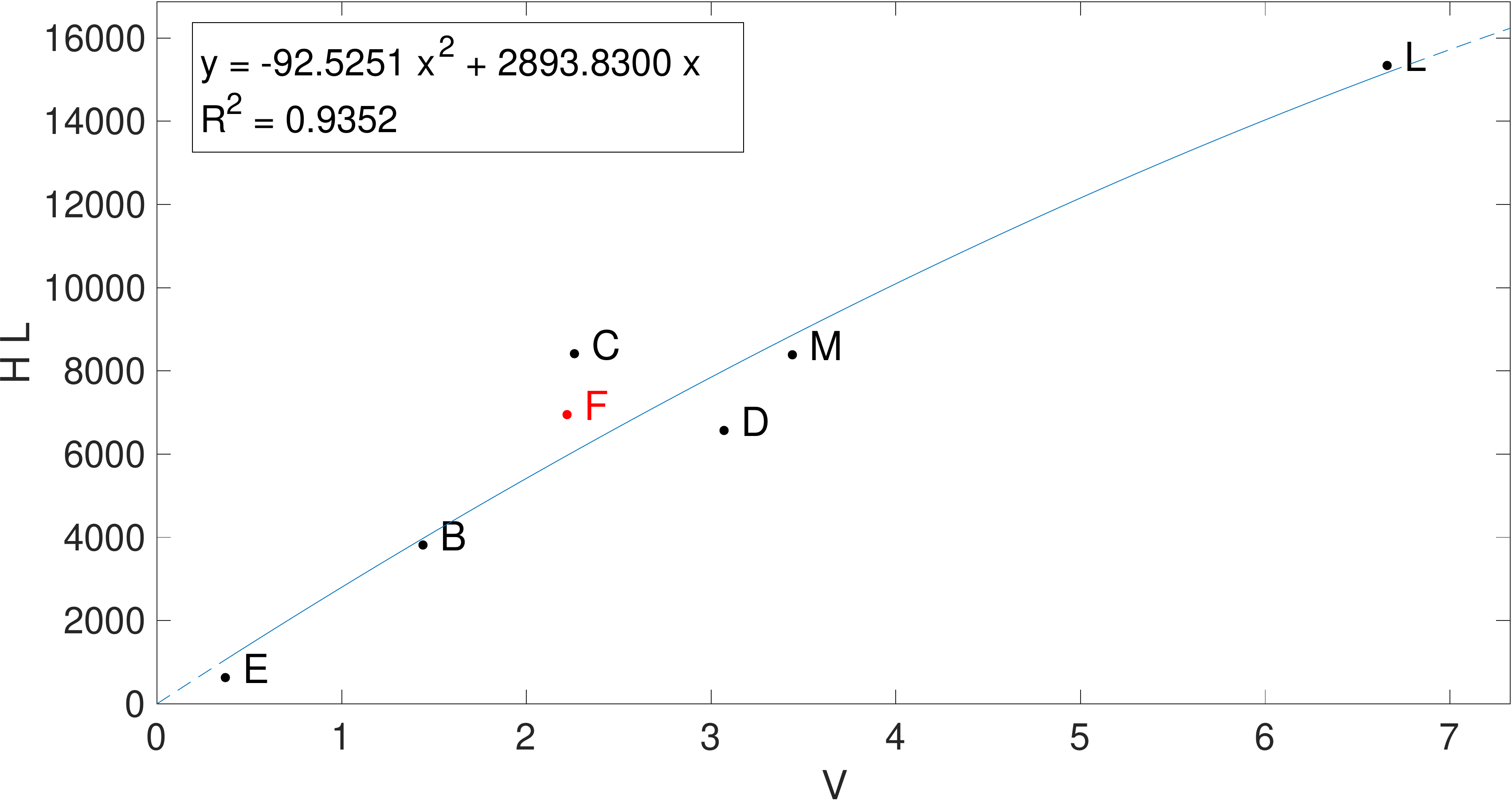}
\caption{Correlation between the landslide volume by the slope and the wave height induced (left panel). Correlation between the landslide volume and the product of the wave height and the wavelength induced (right panel). Outliers in red.}
\label{fig:correlation3}
\end{figure}

Finally and most importantly, the missing LGW characteristics, wave height ($H$) and wavelength ($L$), need to be estimated too.
Again, a number of variable combinations were tested looking for reasonable correlations based on physical insights.
In that sense, \citer{panizzo05} indicate that both in their experiments and in \citer{kamphuis70}, ``greater wave heights are expected at smaller slope inclination angles''.
This means that the wave height should be proportional to the landslide volume and inversely proportional to the local slope (obtained from the bathymetry data shown in Figure~\ref{fig:map}), as we find in the left panel in Figure~\ref{fig:correlation3}.
Based on observation (left panel), the largest wave height (landslide F) has been deemed as an outlier and left out of the fitting, which ends up having a high correlation with the data ($R^2 = 0.8548$).

In order to estimate the wavelength, the product between wave height and wavelength correlates extremely well (right panel, $R^2 = 0.9352$) with the landslide volume.
Since $H$ can be calculated already thanks to the expression in the left panel, $L$ can also be calculated, and both variables are included in blue font in Table~\ref{tab:landslideCharacteristics}.

It is important to remark that the new estimated values for all variables lie within the bounds of existing landslides.
Therefore, the simple models developed are not being applied out of their range and used to interpolate rather than to extrapolate the magnitudes.

\subsection{Landslide-generated tsunami wave propagation}
\label{sec:propagation}

The tsunami wave propagation has been simulated using the NLSW model COMCOT \cite{wang09}.
The numerical setup is the same used in \citer{liu20}, namely with a bathymetry derived from the BIG14 dataset, with a resolution of $~20$ m (0.0012 arc-minutes) and a constant bottom friction Manning coefficient of 0.013 \cite{garzon16}.
The simulation time is 20 minutes, enough to observe the development and evolution of the largest tsunami waves, and requires approximately 5 hours to compute.

Each of the 14 waves has been simulated independently to establish their relative contribution to the waves observed in Palu, Wani and Pantoloan (see Figure~\ref{fig:map}).
As a starting point, the wave characteristics ($H$ and $\lambda$) tested are those included in Table~\ref{tab:landslideCharacteristics}, and the waves have been considered to start propagating perpendicularly to the shoreline where they were generated.

\begin{figure}
\centering
\includegraphics[width=0.19\textwidth]{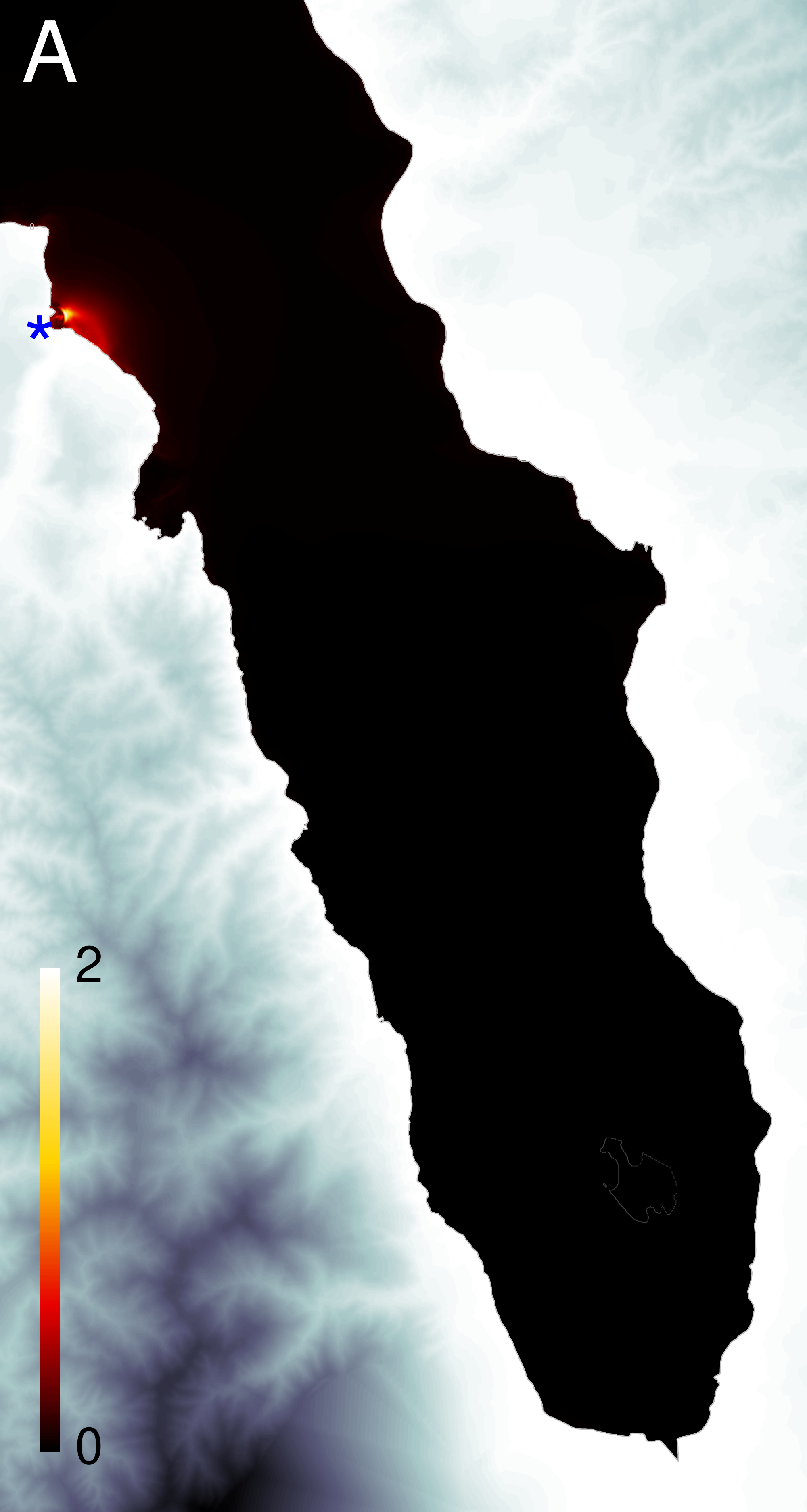}
\includegraphics[width=0.19\textwidth]{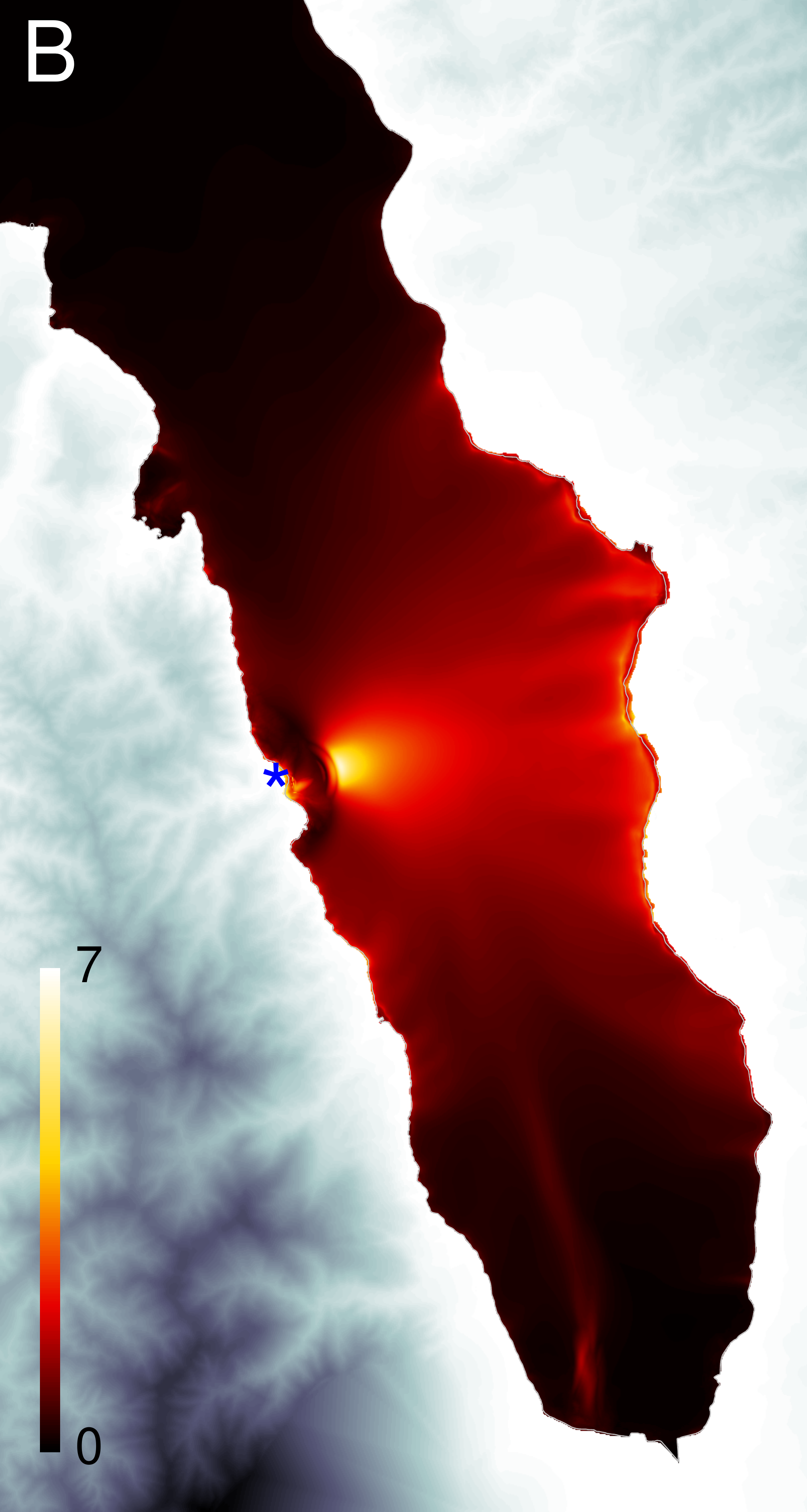}
\includegraphics[width=0.19\textwidth]{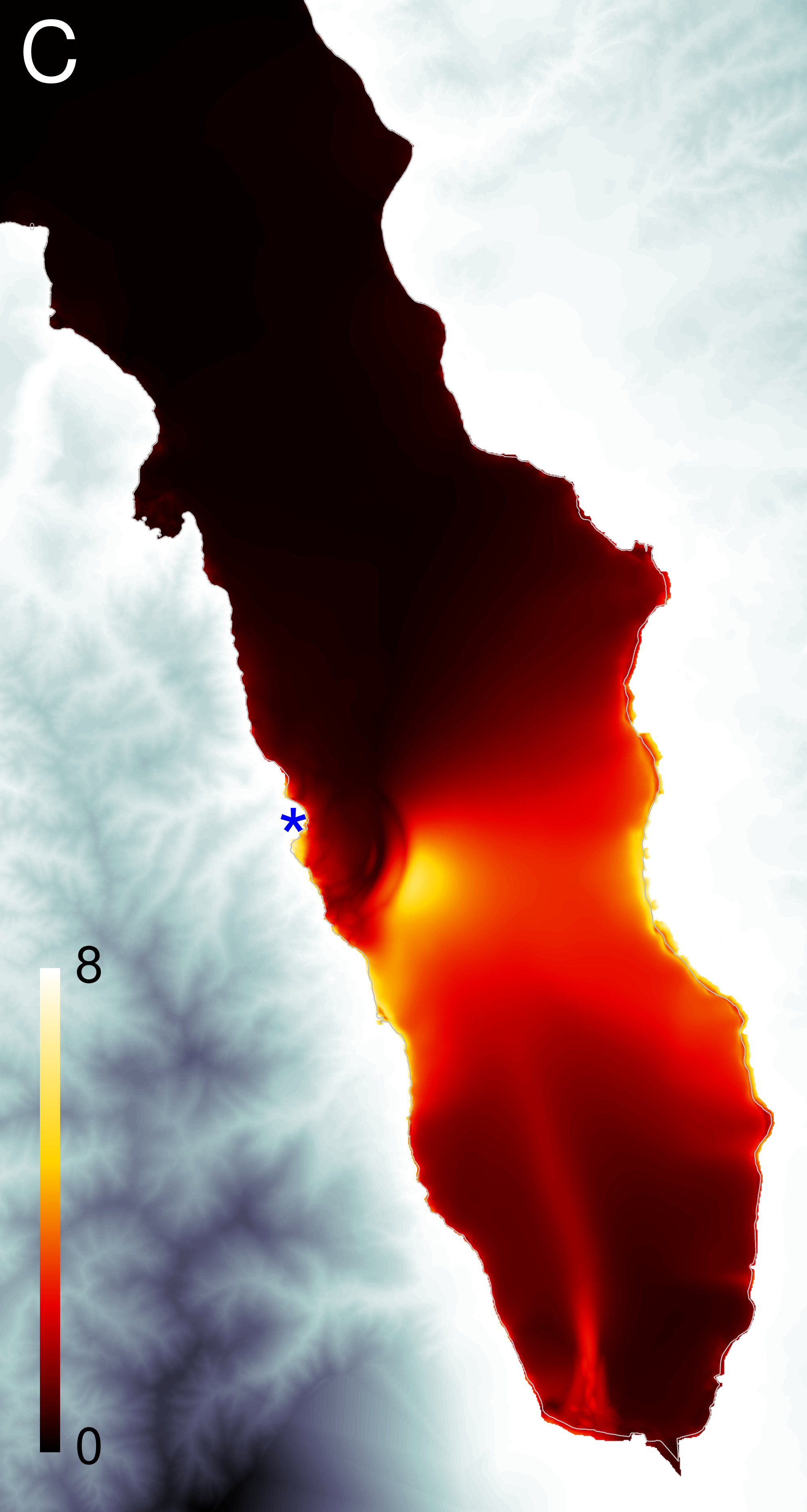}
\includegraphics[width=0.19\textwidth]{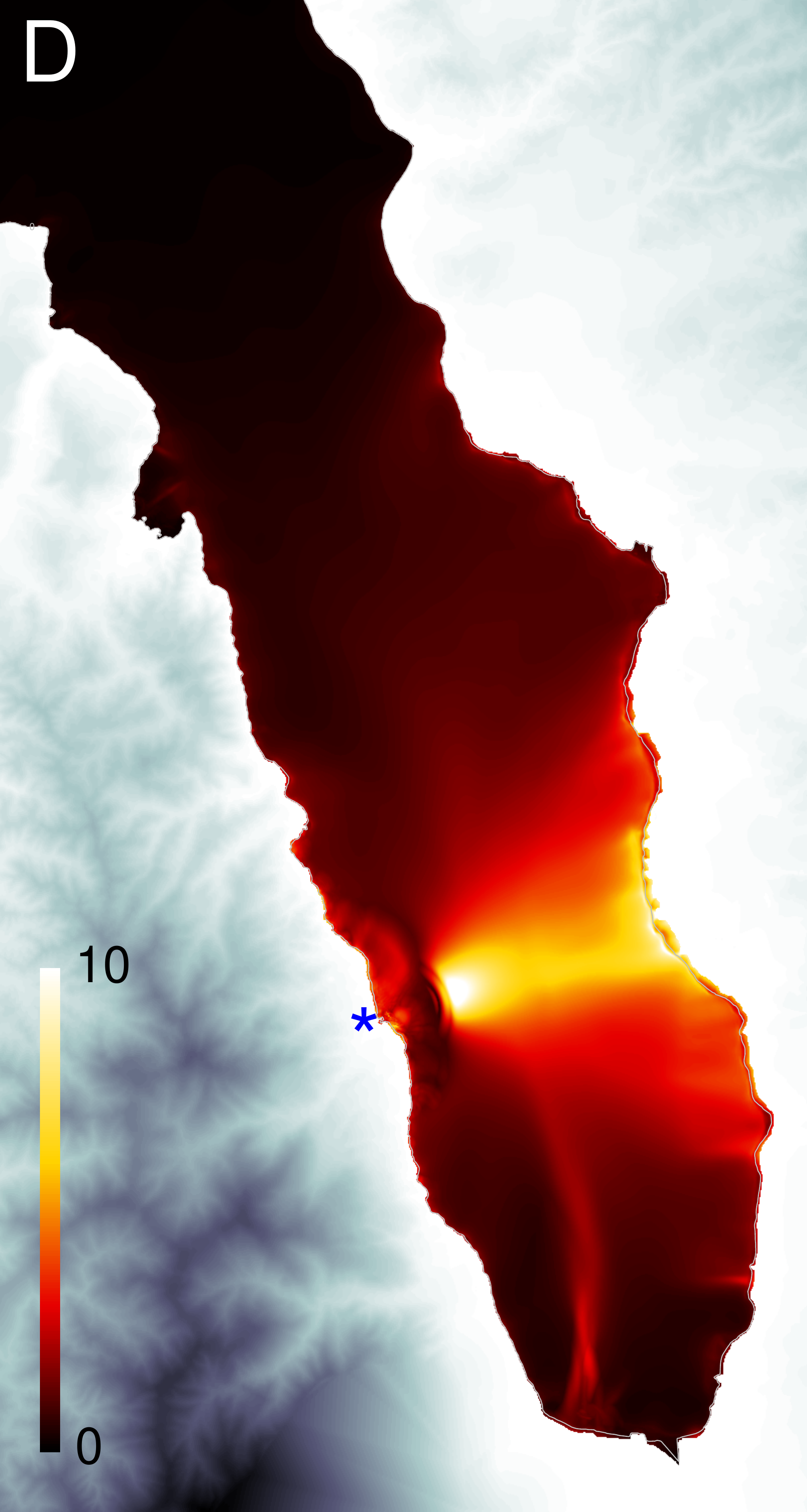}
\includegraphics[width=0.19\textwidth]{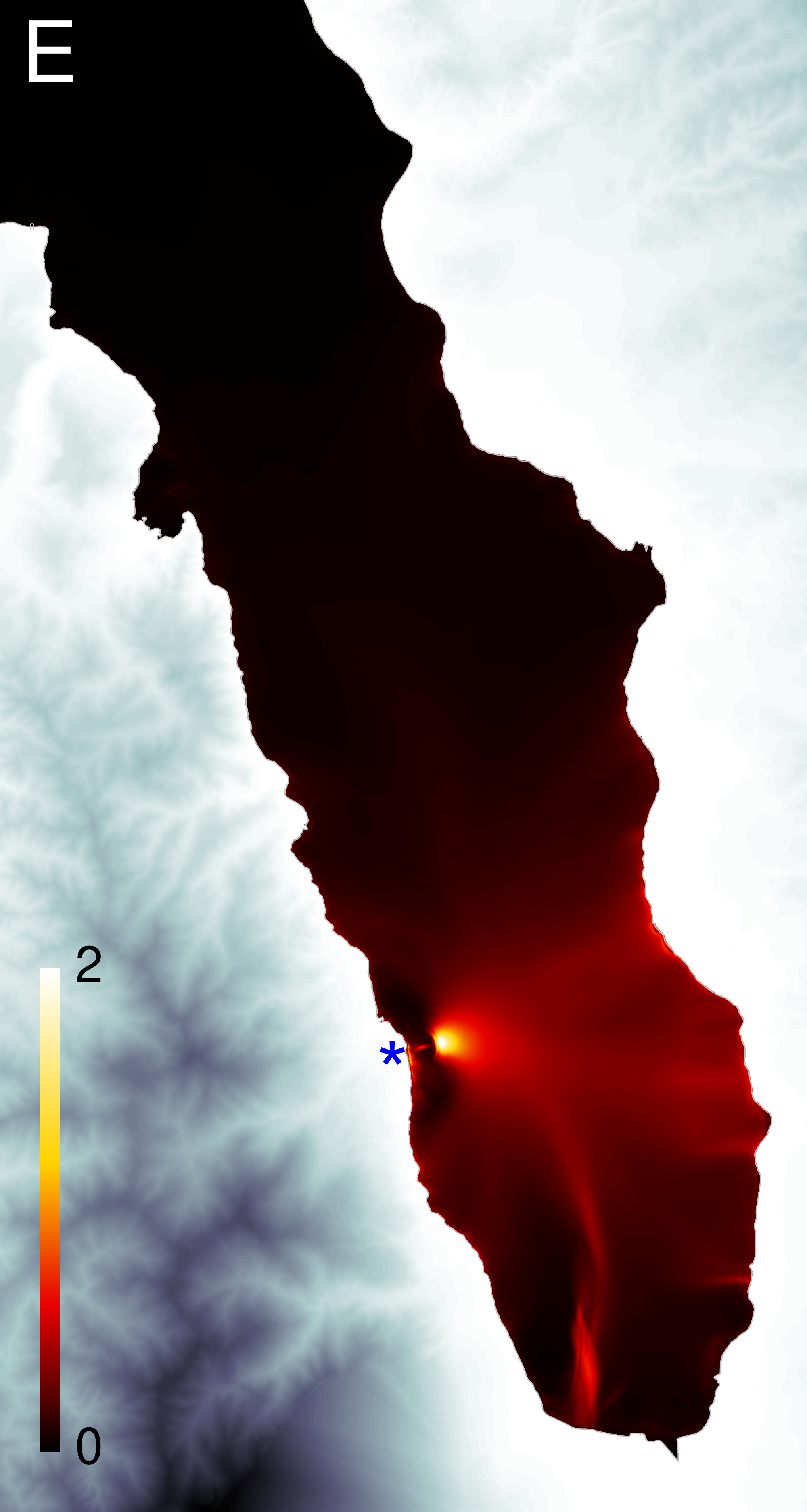}\\
\includegraphics[width=0.19\textwidth]{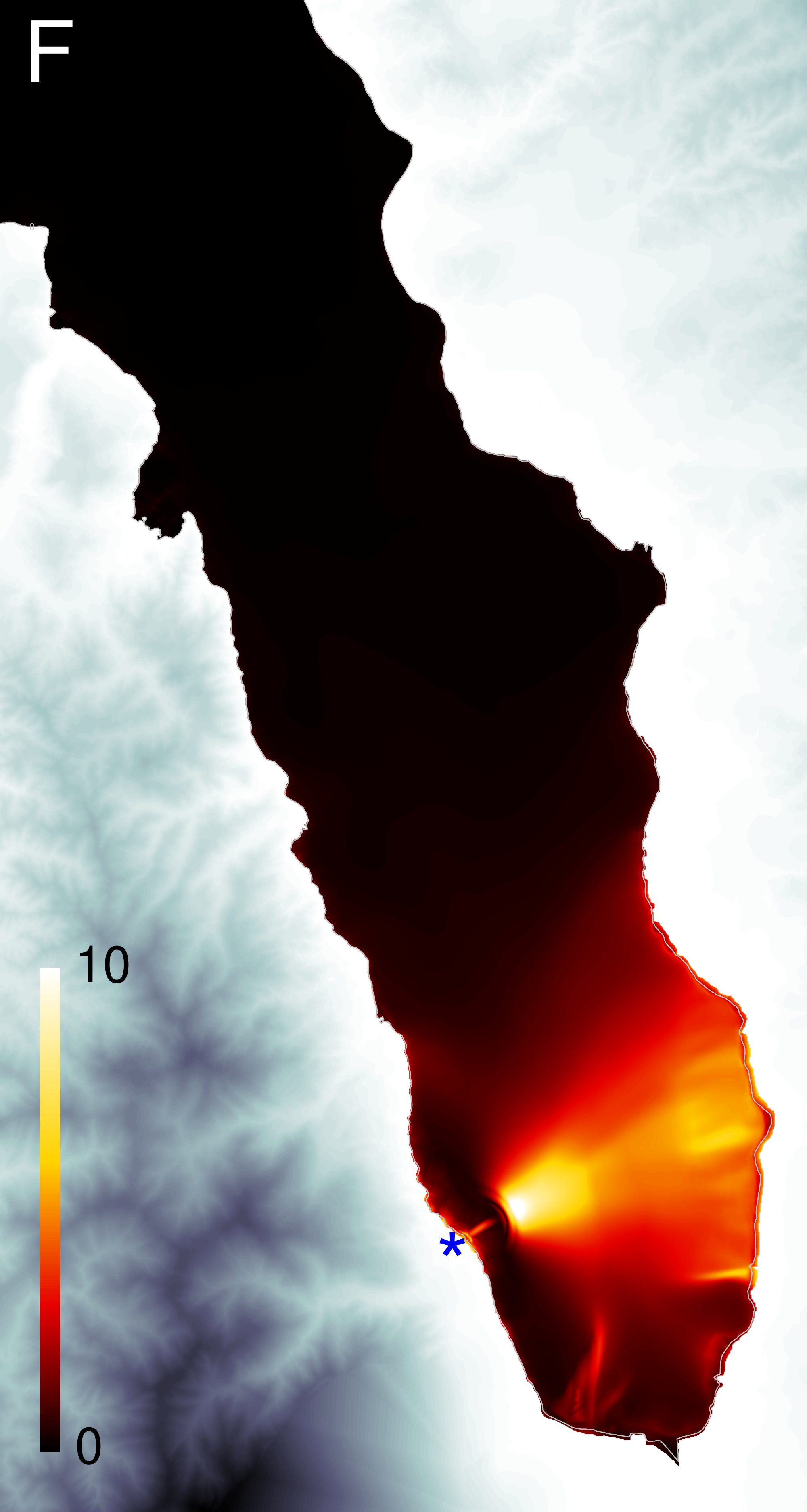}
\includegraphics[width=0.19\textwidth]{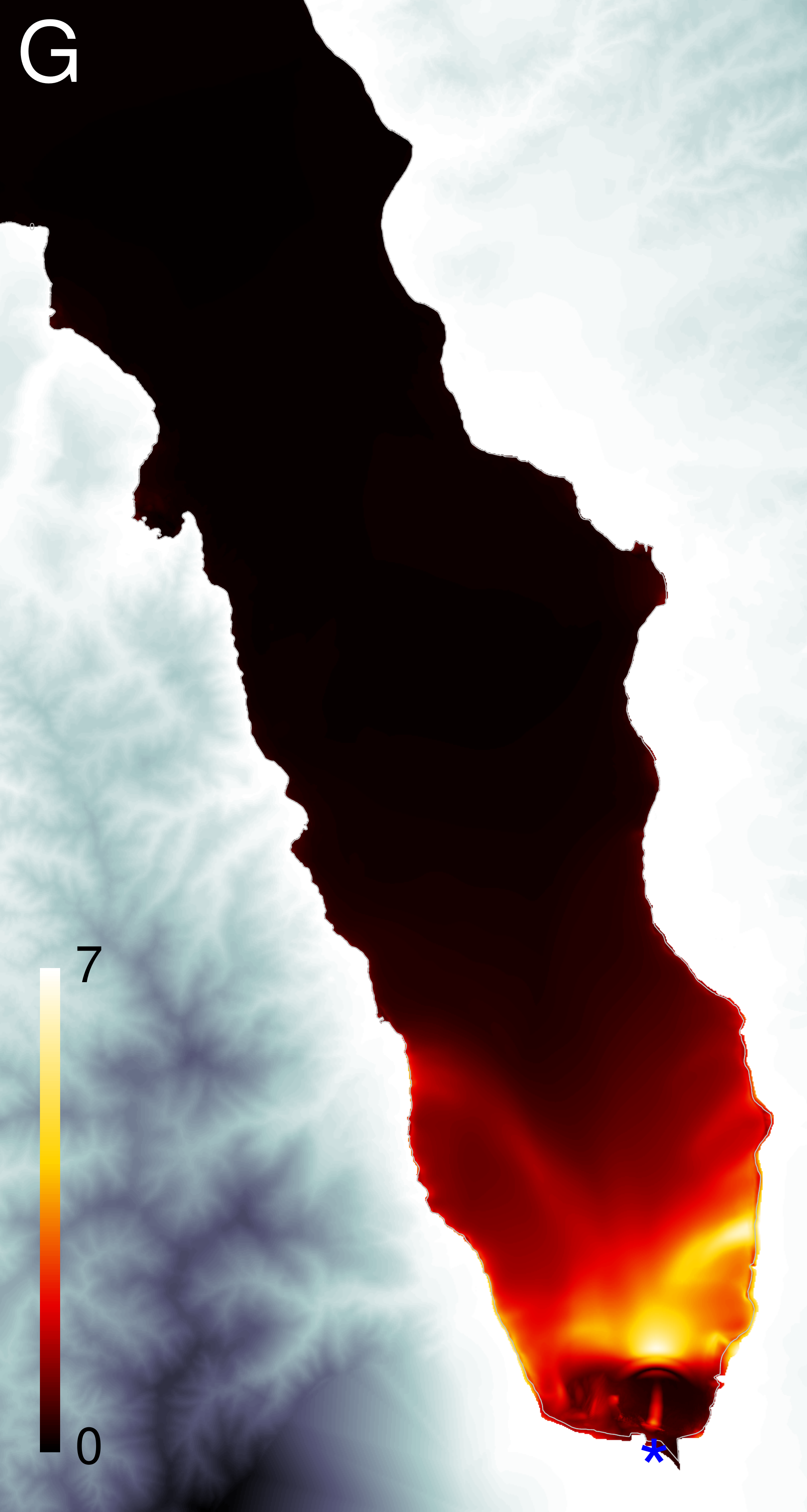}
\includegraphics[width=0.19\textwidth]{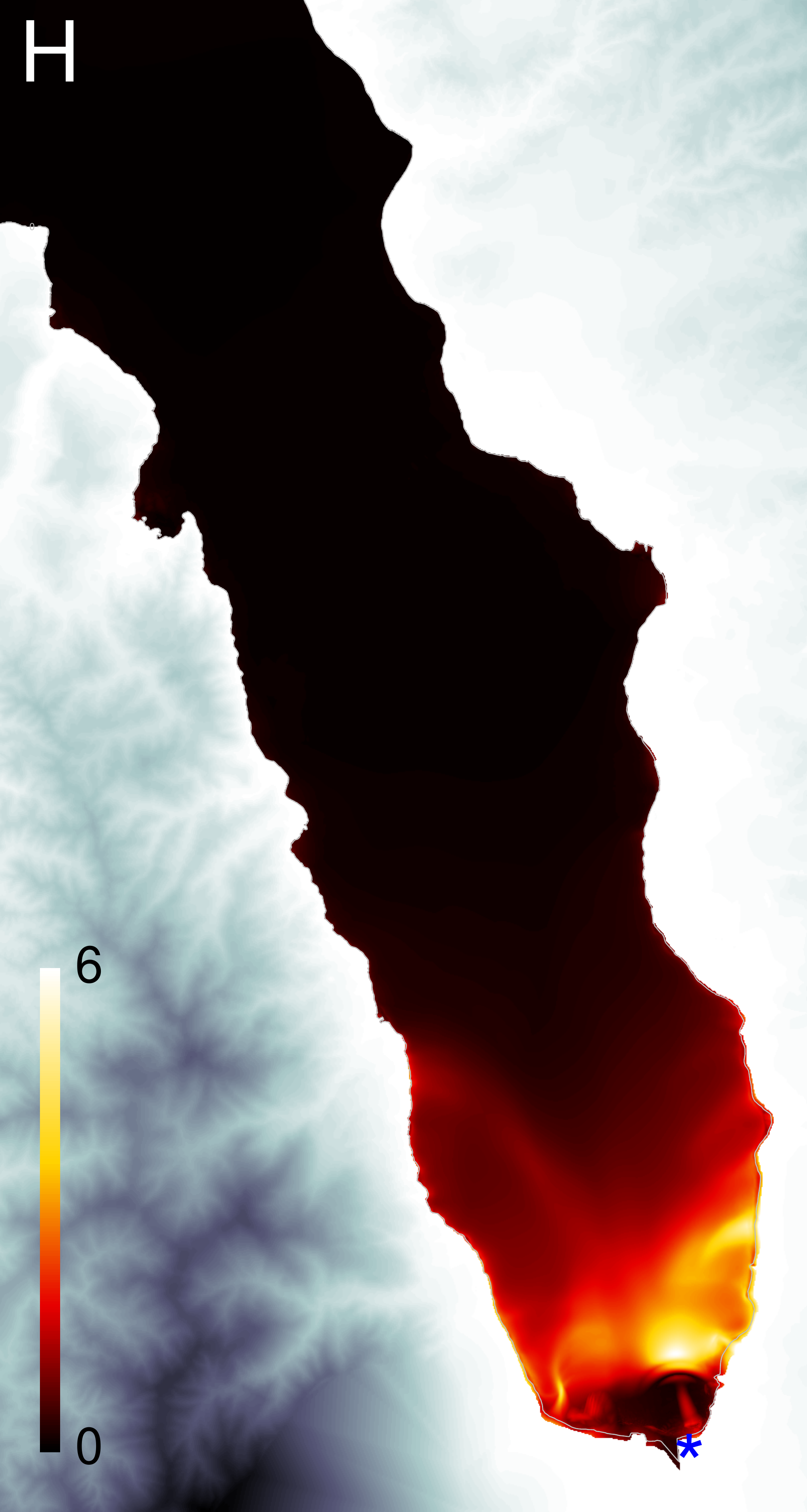}
\includegraphics[width=0.19\textwidth]{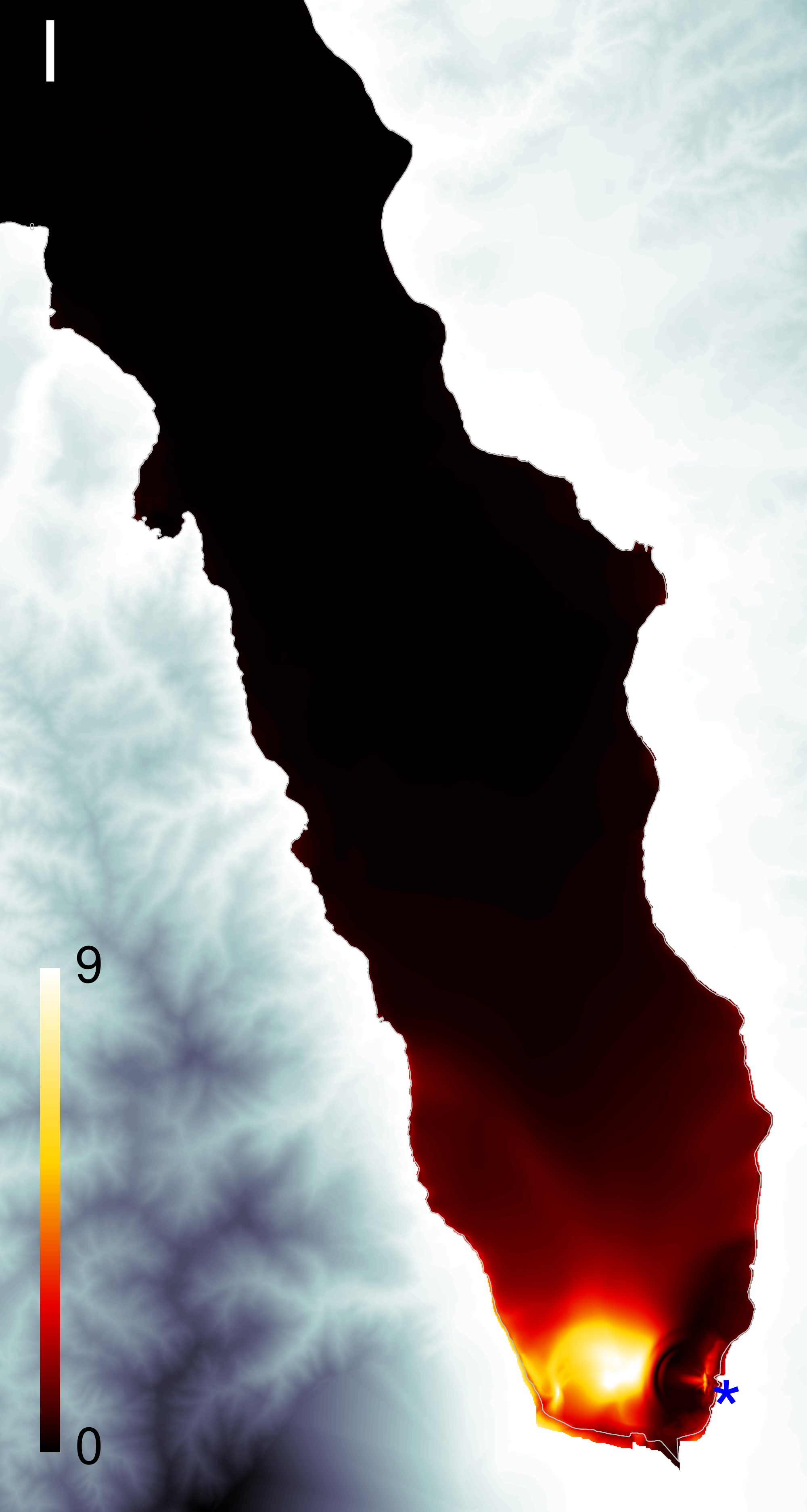}
\includegraphics[width=0.19\textwidth]{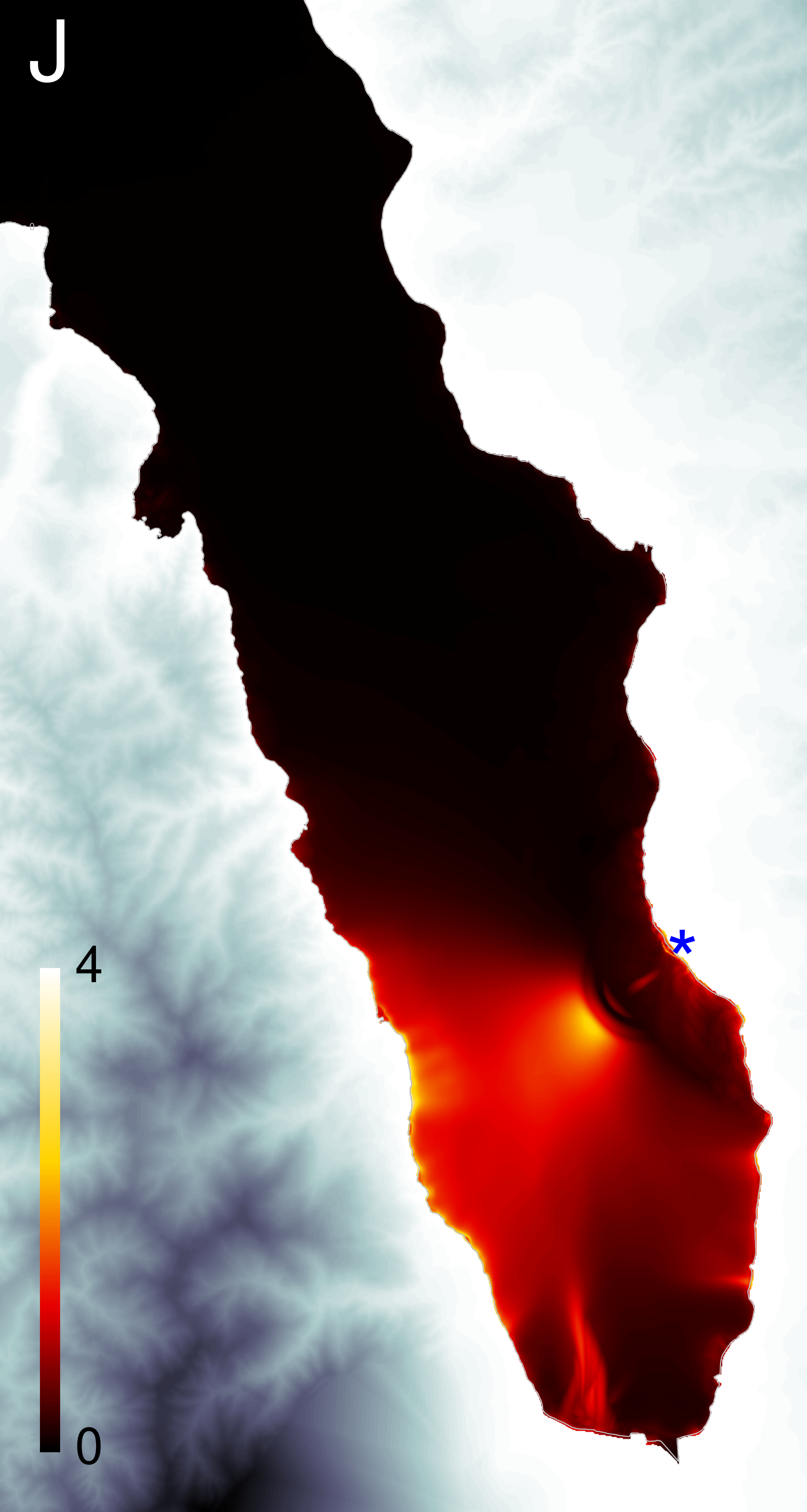}\\
\includegraphics[width=0.19\textwidth]{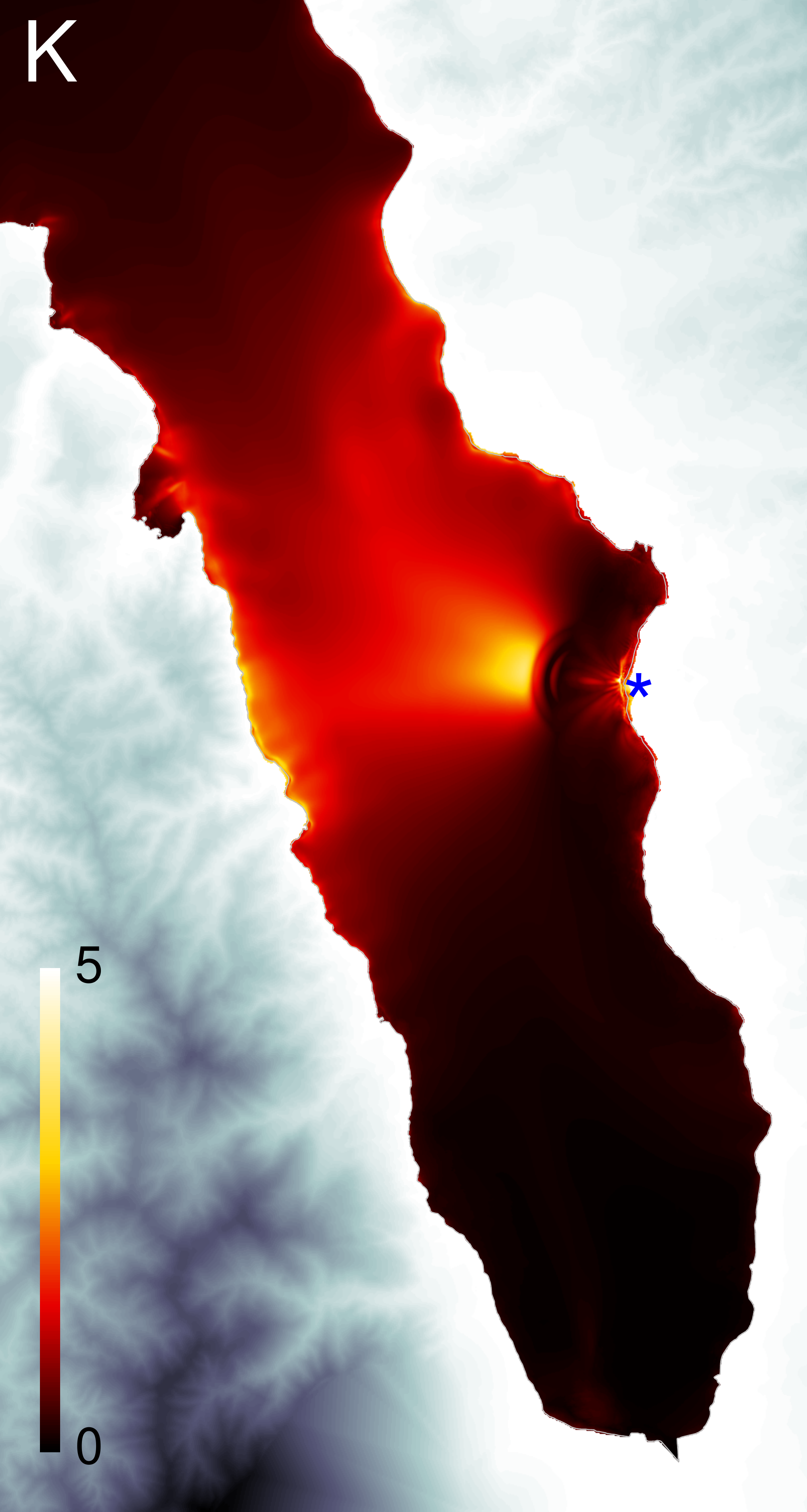}
\includegraphics[width=0.19\textwidth]{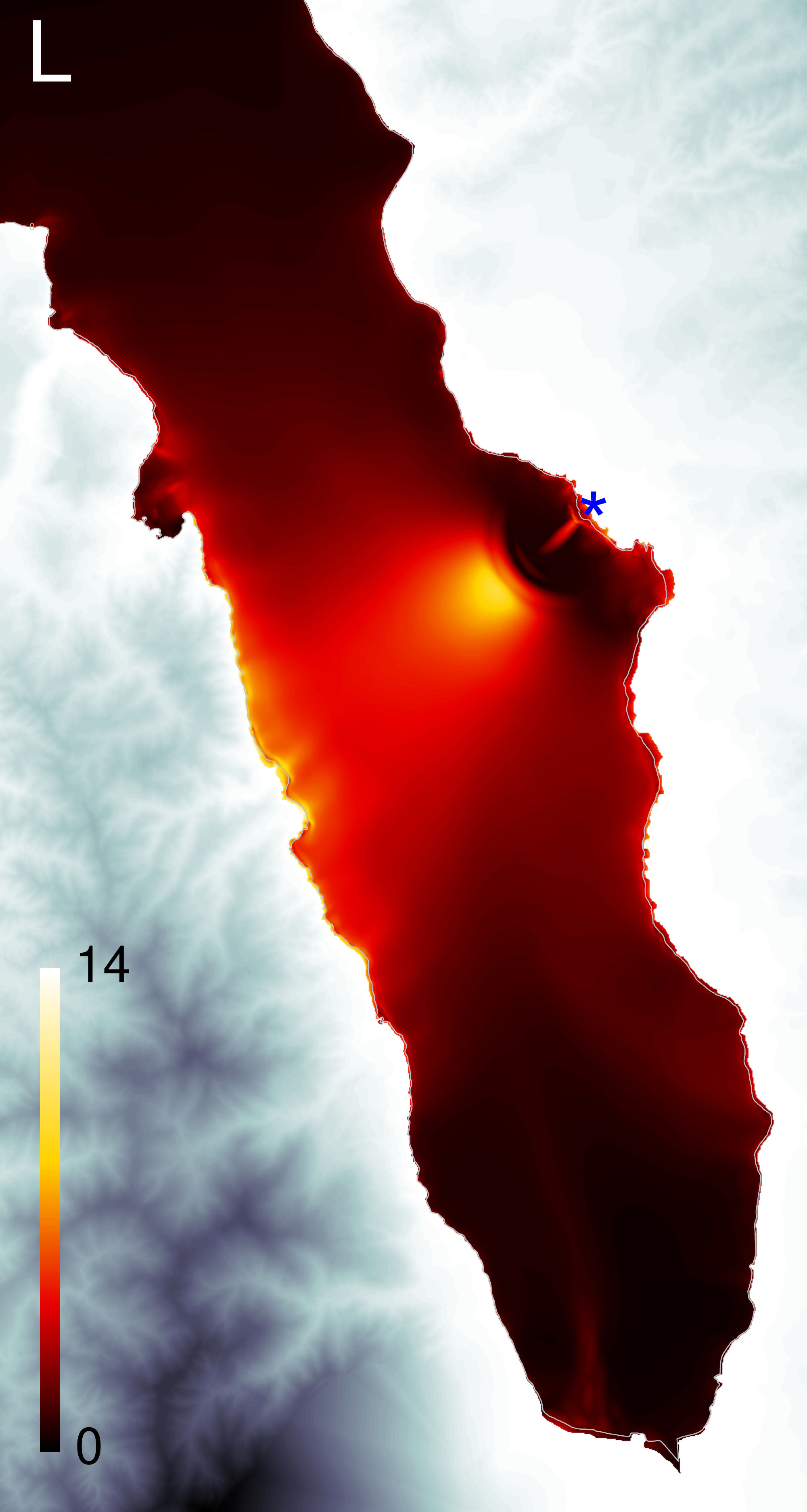}
\includegraphics[width=0.19\textwidth]{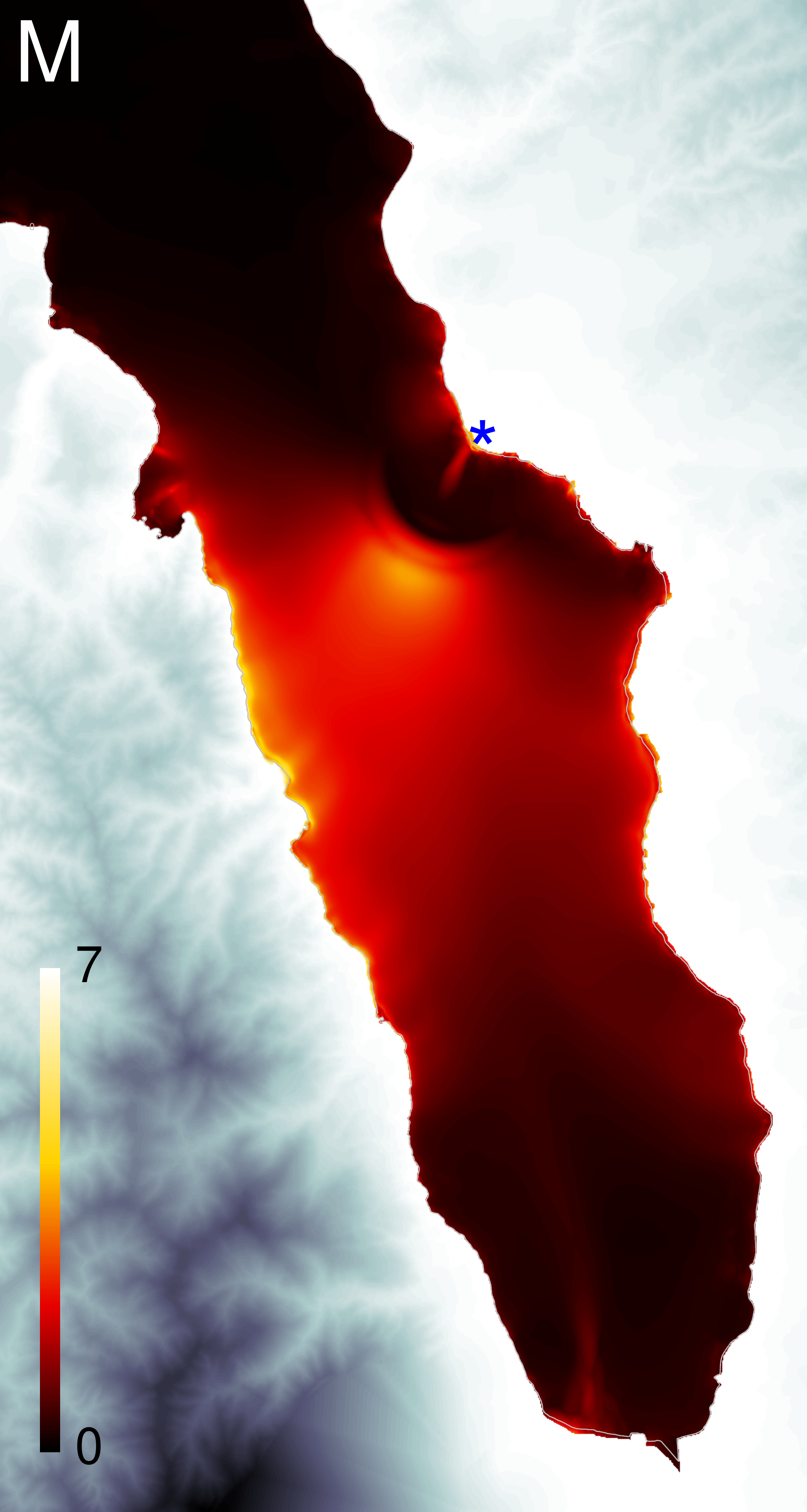}
\includegraphics[width=0.19\textwidth]{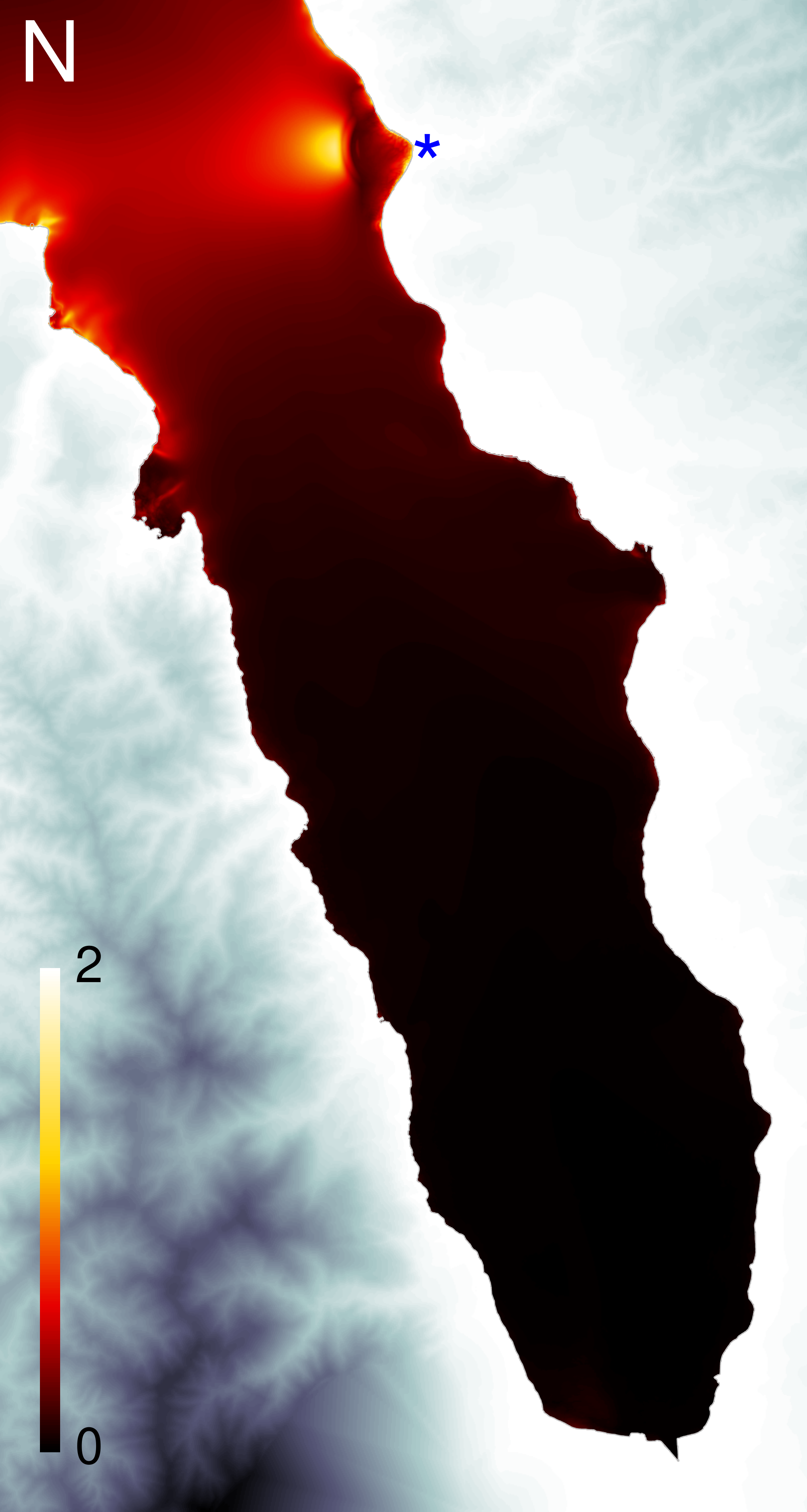}
\caption{Highest free surface elevation induced independently by each of the landslides in Palu Bay. Colorbar scale in metres. Landslide location denoted by a blue star.}
\label{fig:maxEta}
\end{figure}

\begin{figure}
\centering
\includegraphics[trim=0 35 0 0,clip,width=0.49\textwidth]{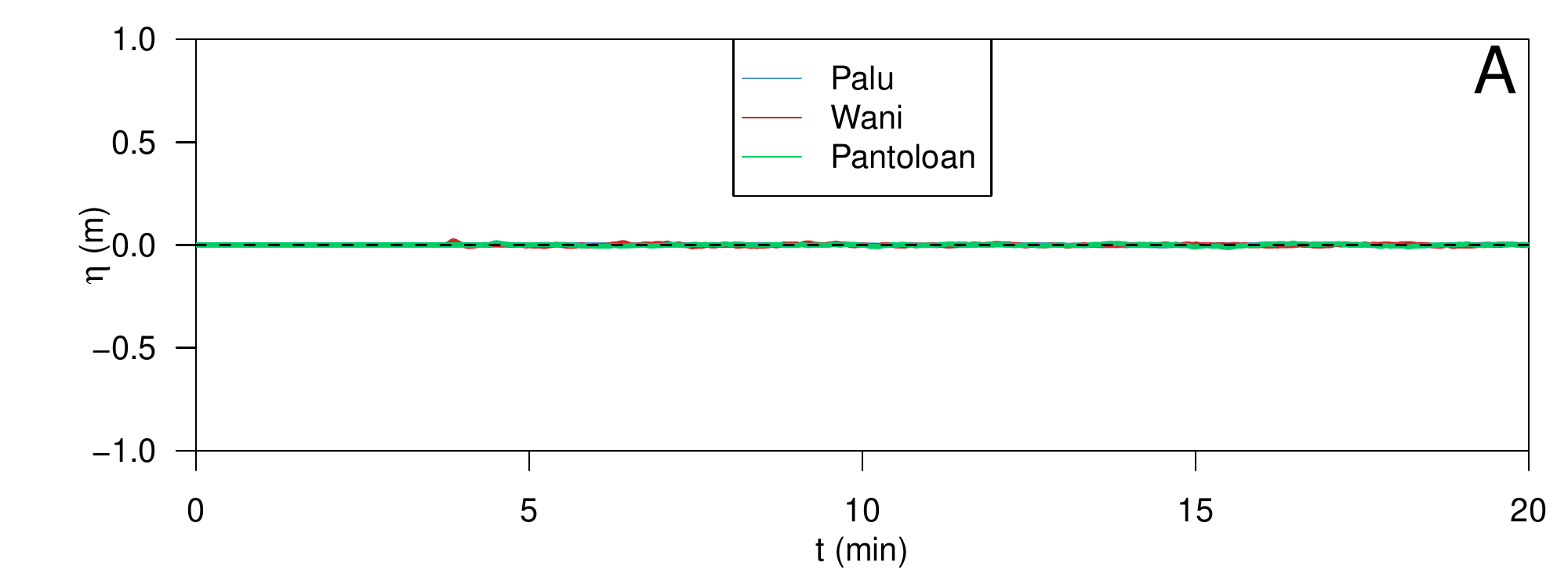}
\includegraphics[trim=0 35 0 0,clip,width=0.49\textwidth]{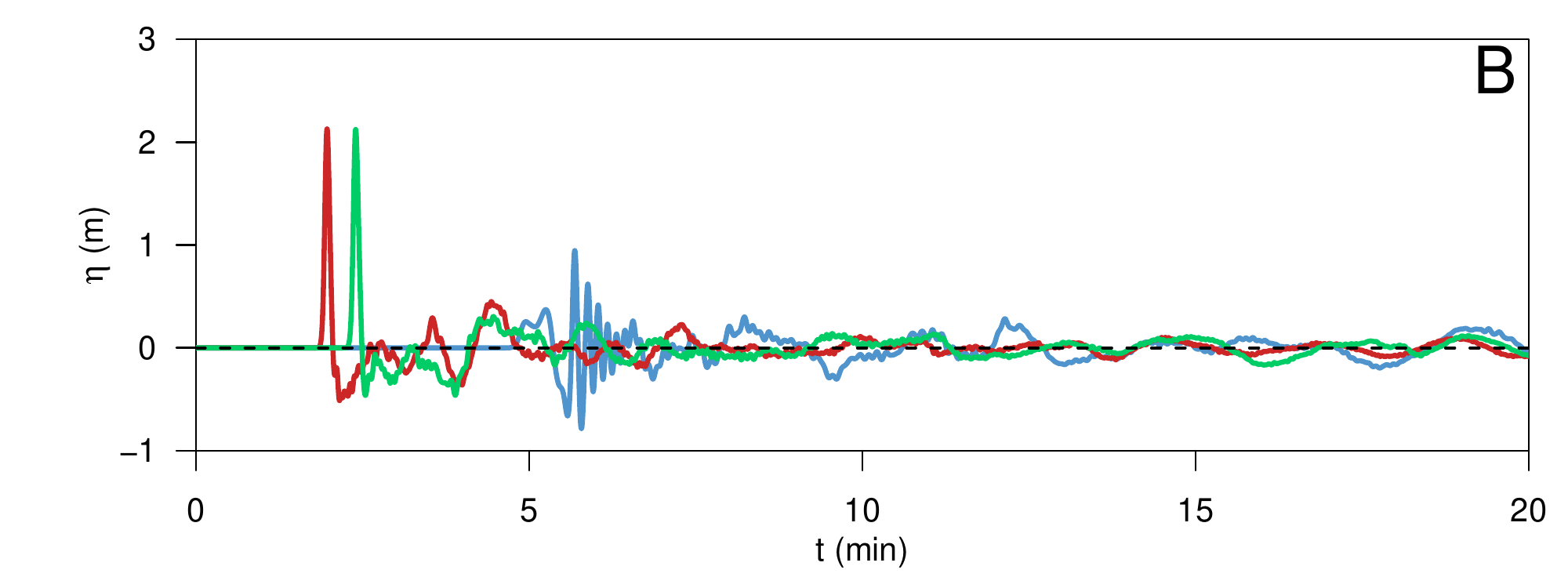}\\
\includegraphics[trim=0 35 0 0,clip,width=0.49\textwidth]{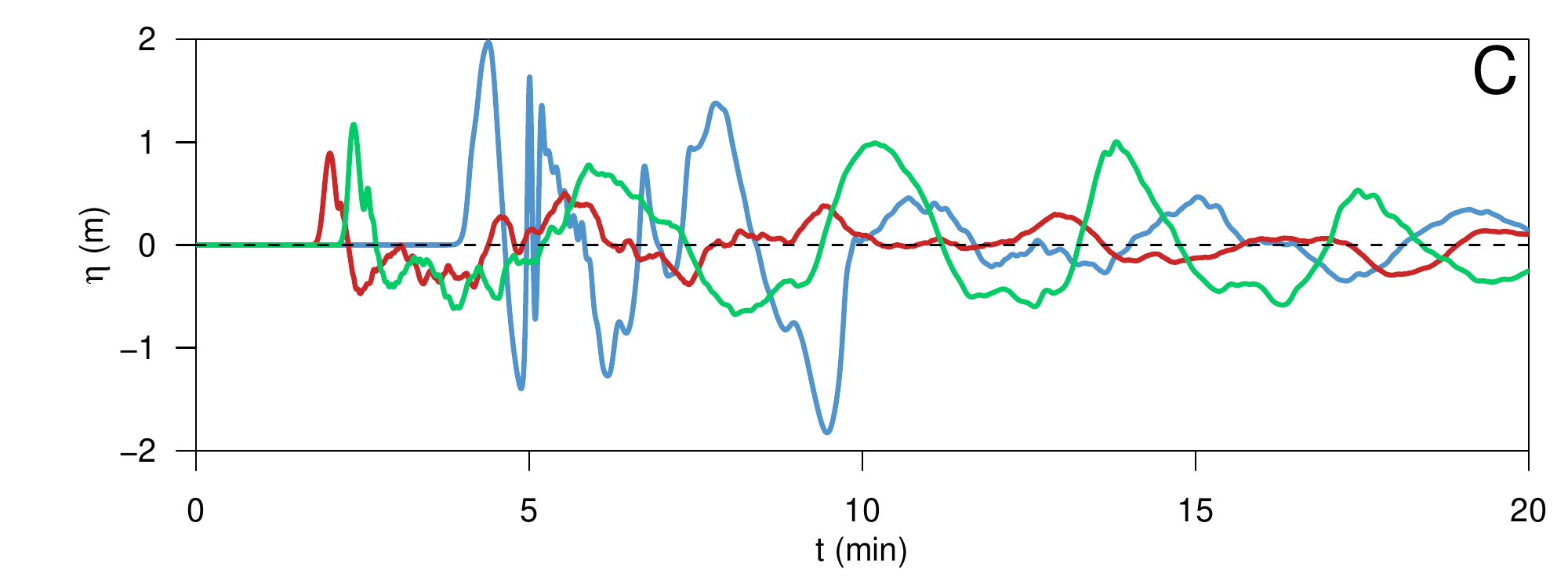}
\includegraphics[trim=0 35 0 0,clip,width=0.49\textwidth]{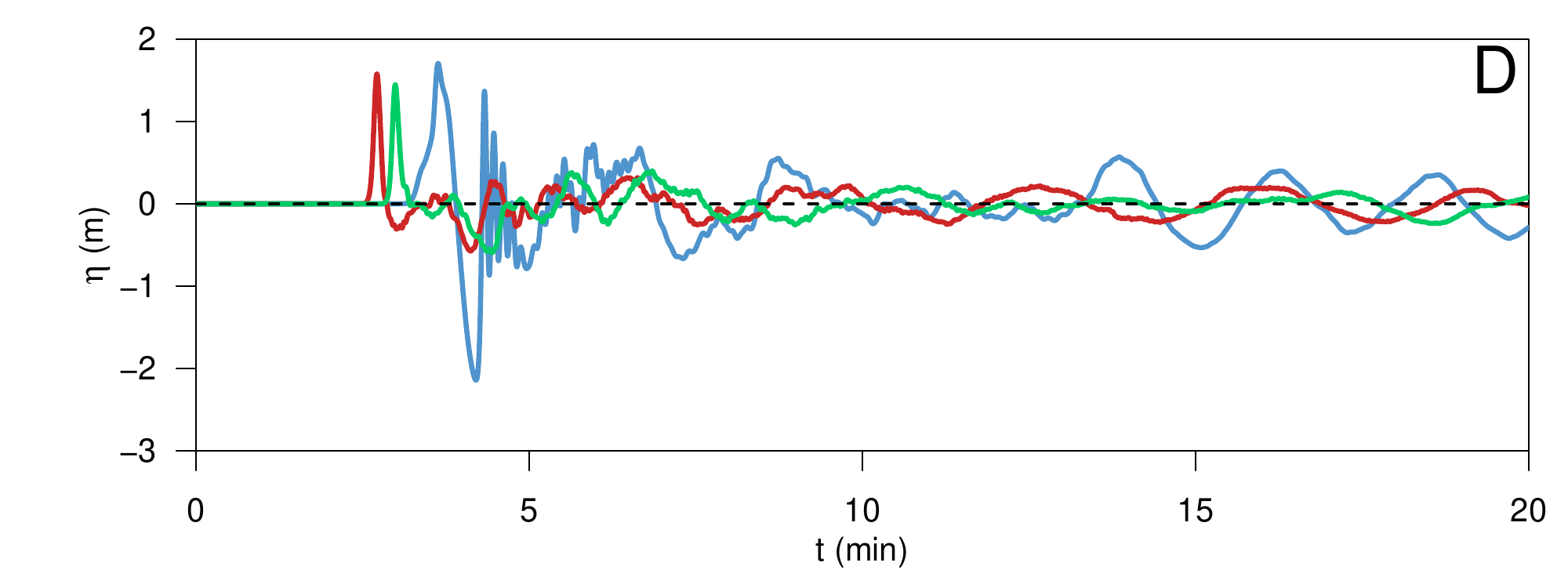}\\
\includegraphics[trim=0 35 0 0,clip,width=0.49\textwidth]{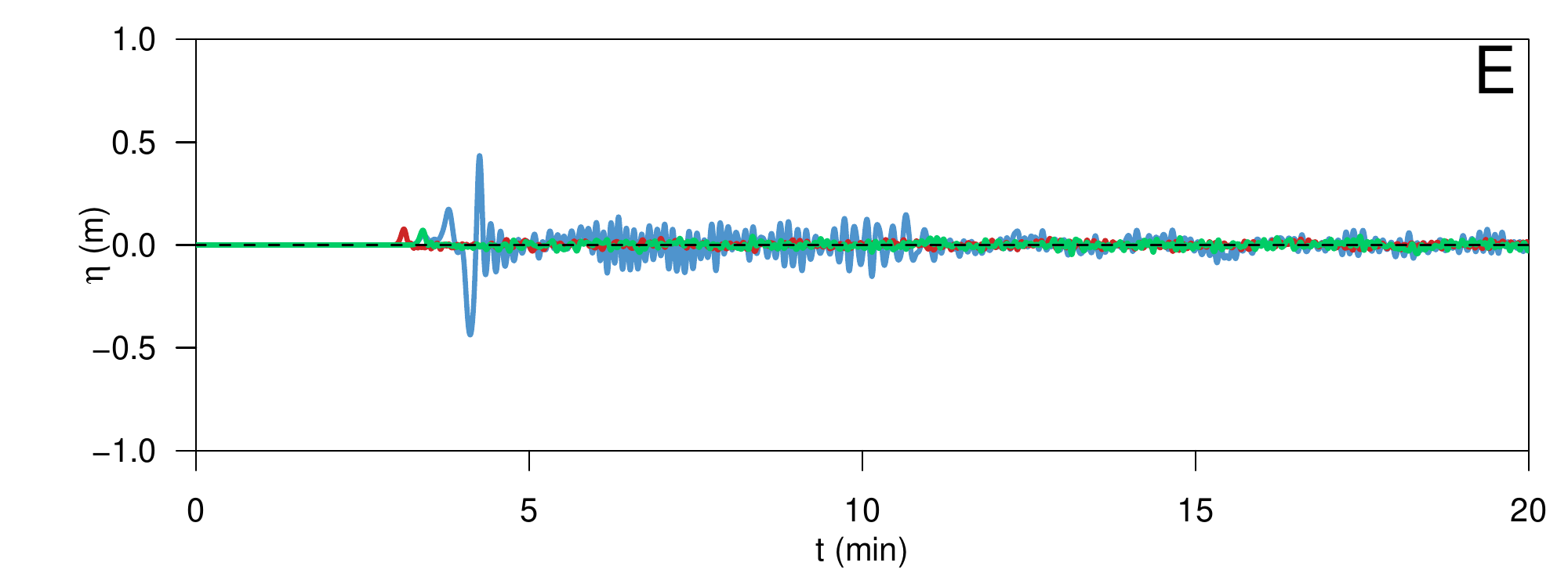}
\includegraphics[trim=0 35 0 0,clip,width=0.49\textwidth]{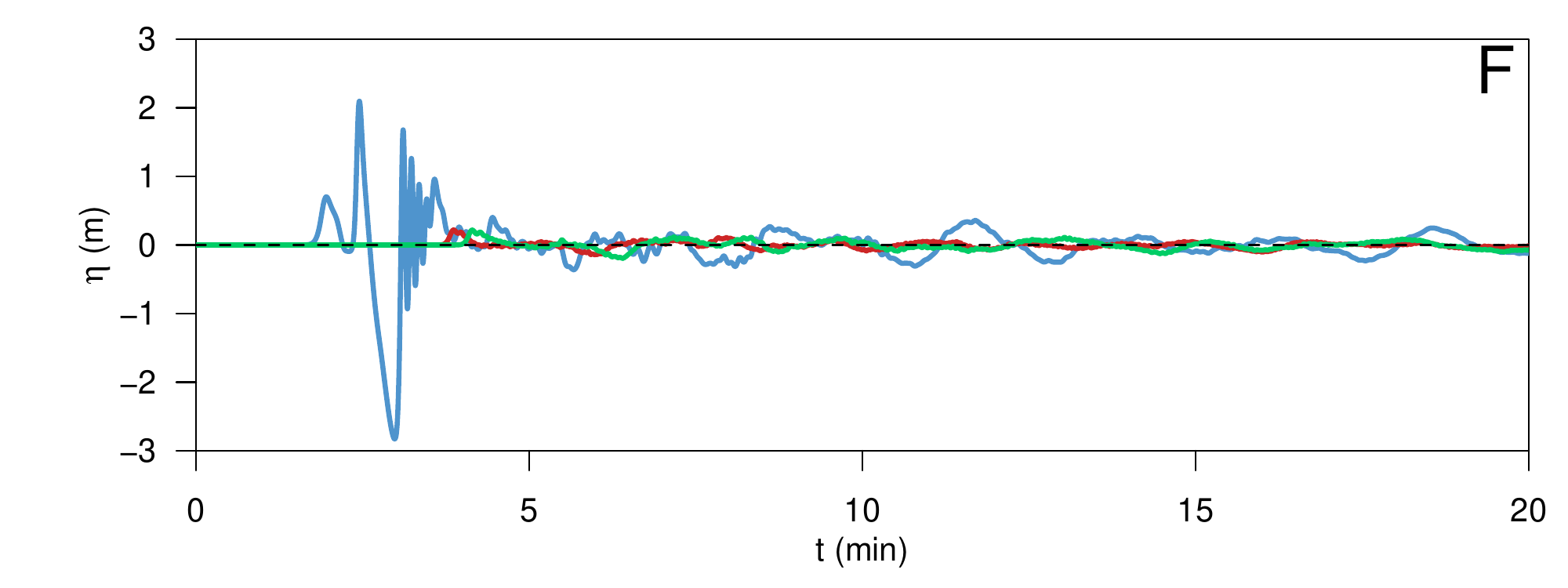}\\
\includegraphics[trim=0 35 0 0,clip,width=0.49\textwidth]{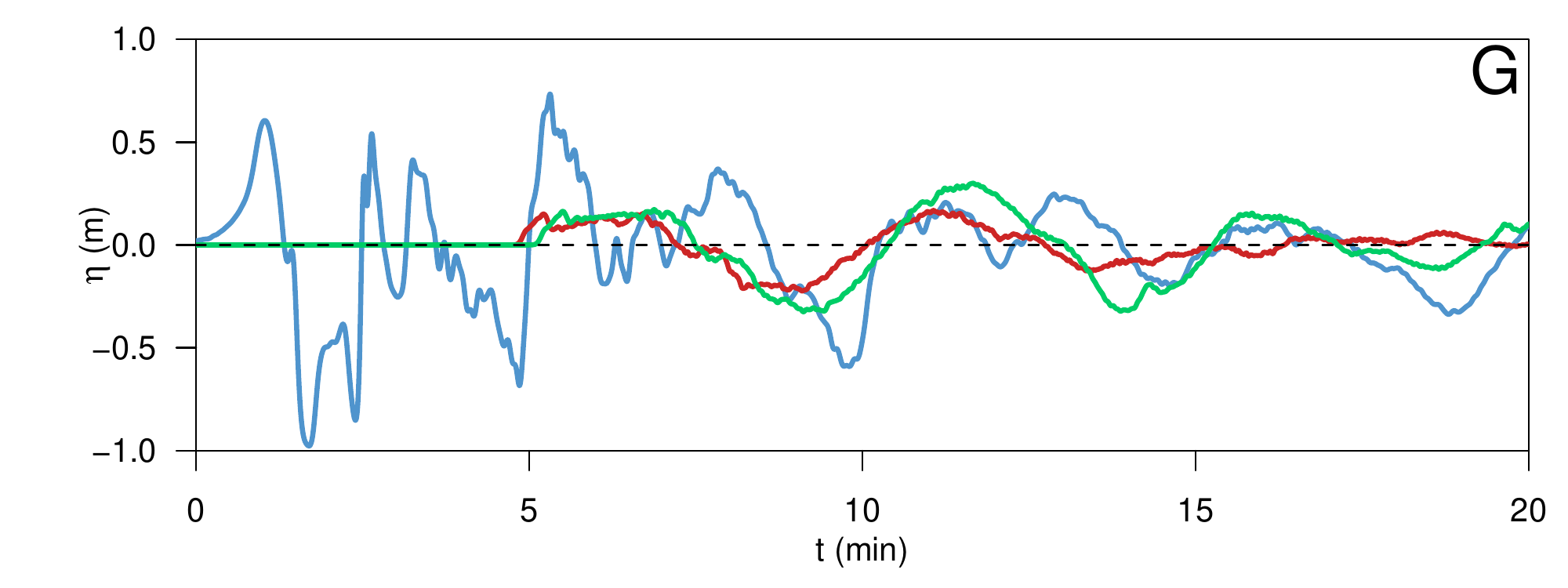}
\includegraphics[trim=0 35 0 0,clip,width=0.49\textwidth]{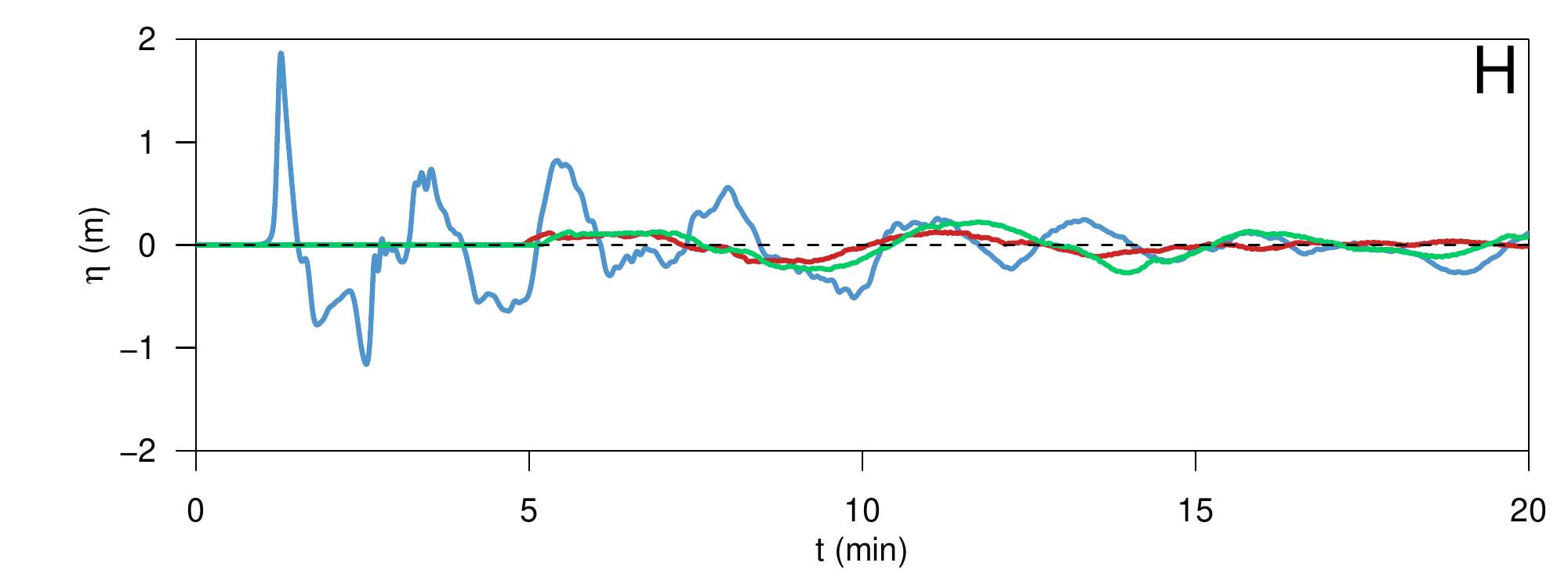}\\
\includegraphics[trim=0 35 0 0,clip,width=0.49\textwidth]{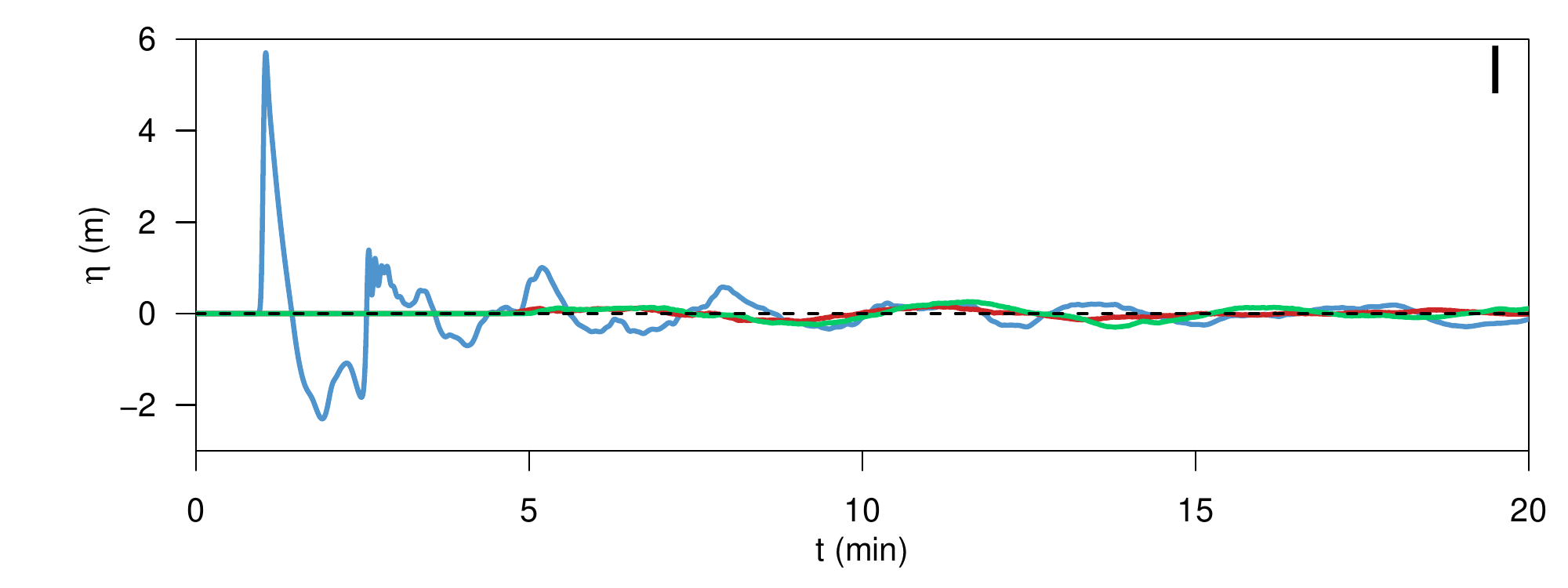}
\includegraphics[trim=0 35 0 0,clip,width=0.49\textwidth]{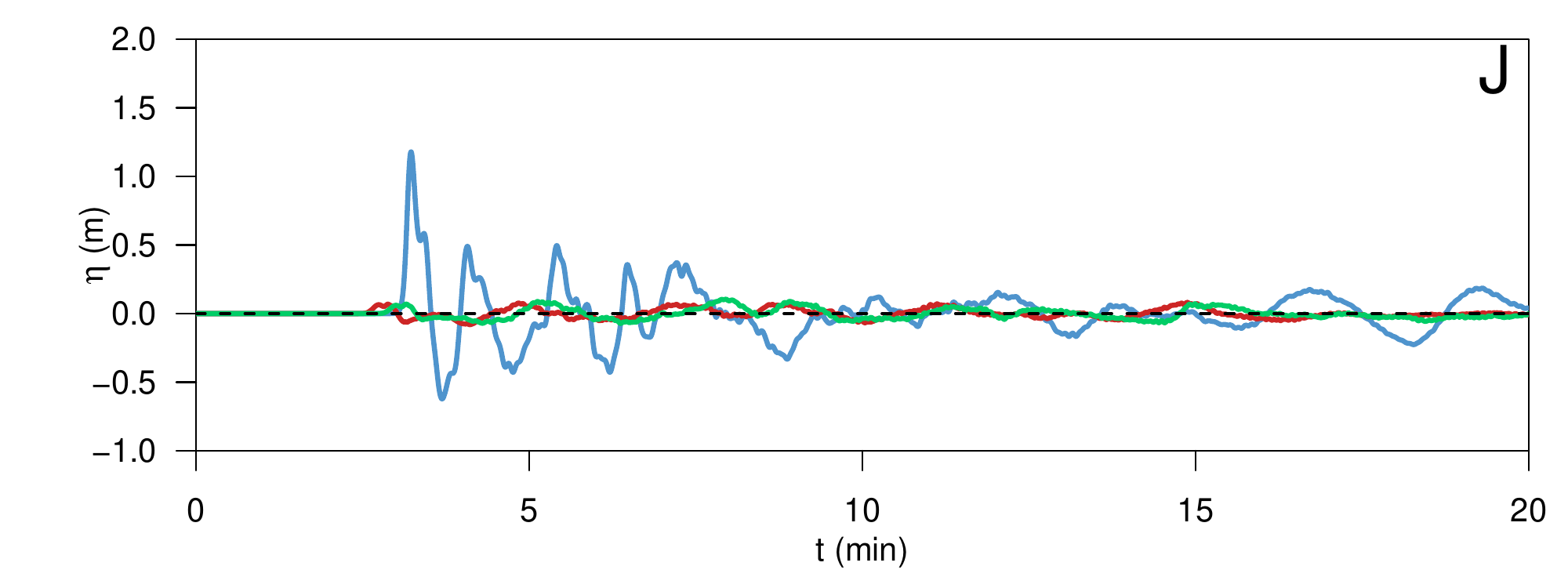}\\
\includegraphics[trim=0 35 0 0,clip,width=0.49\textwidth]{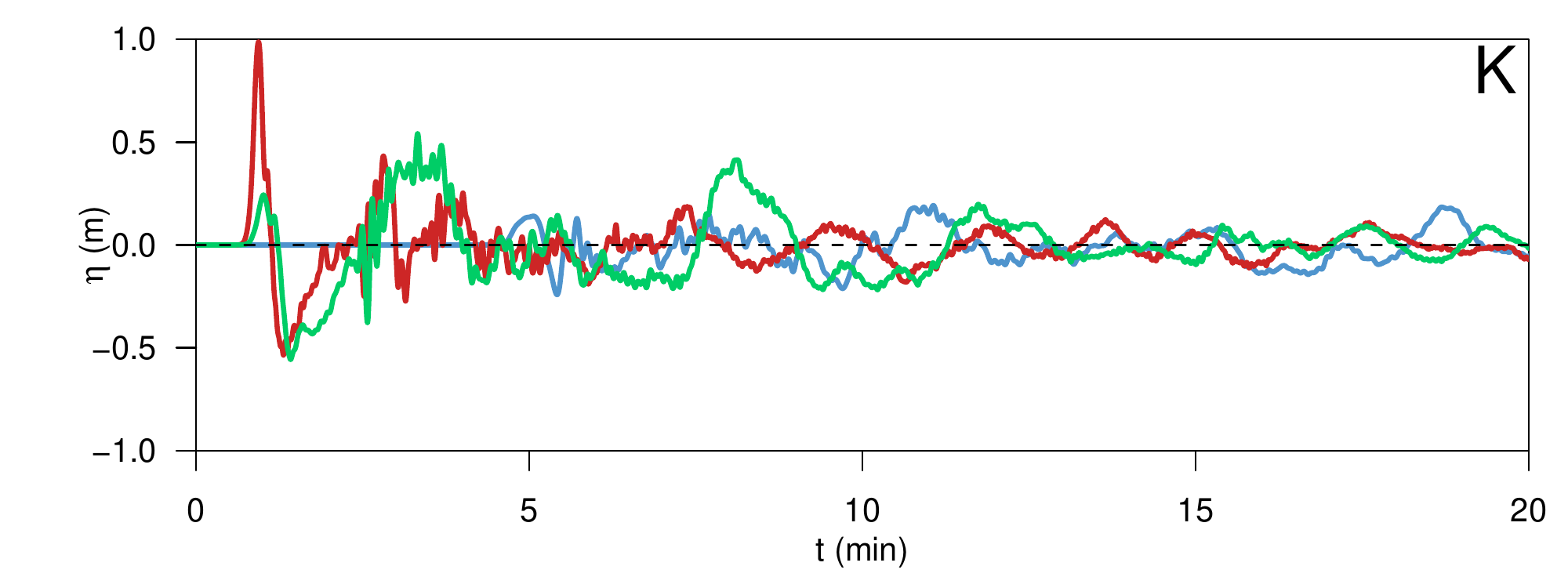}
\includegraphics[trim=0 35 0 0,clip,width=0.49\textwidth]{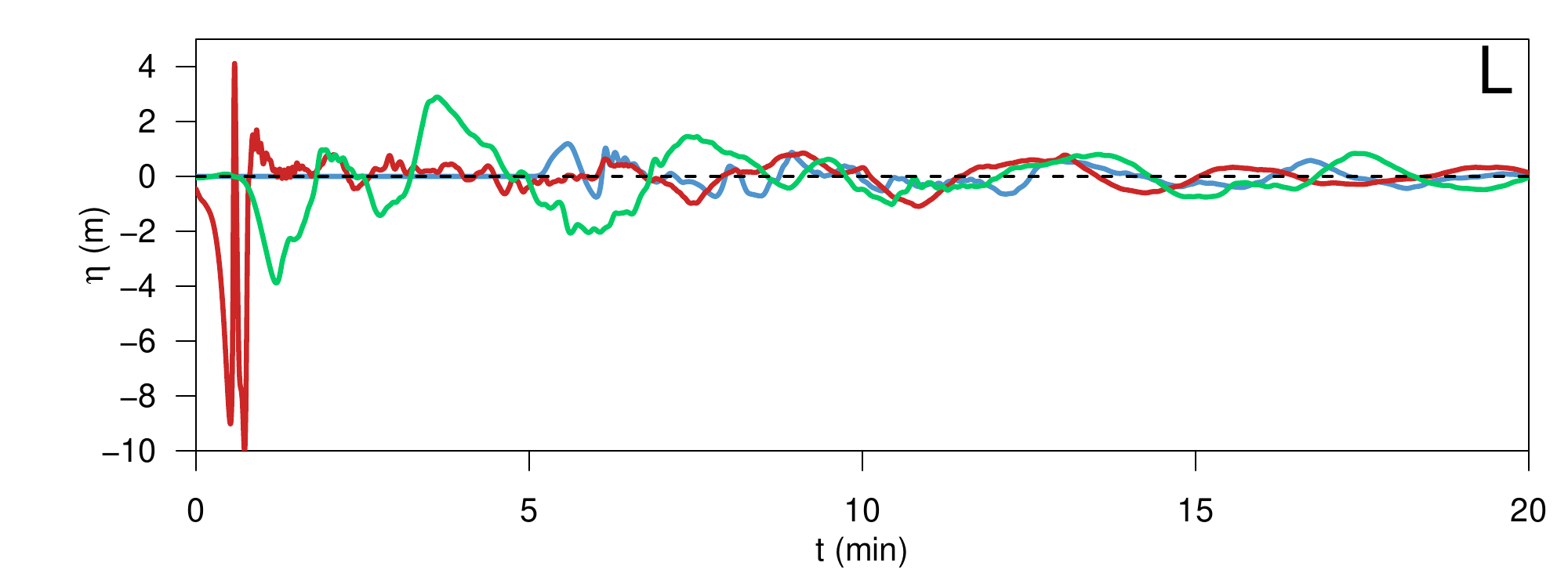}\\
\includegraphics[width=0.49\textwidth]{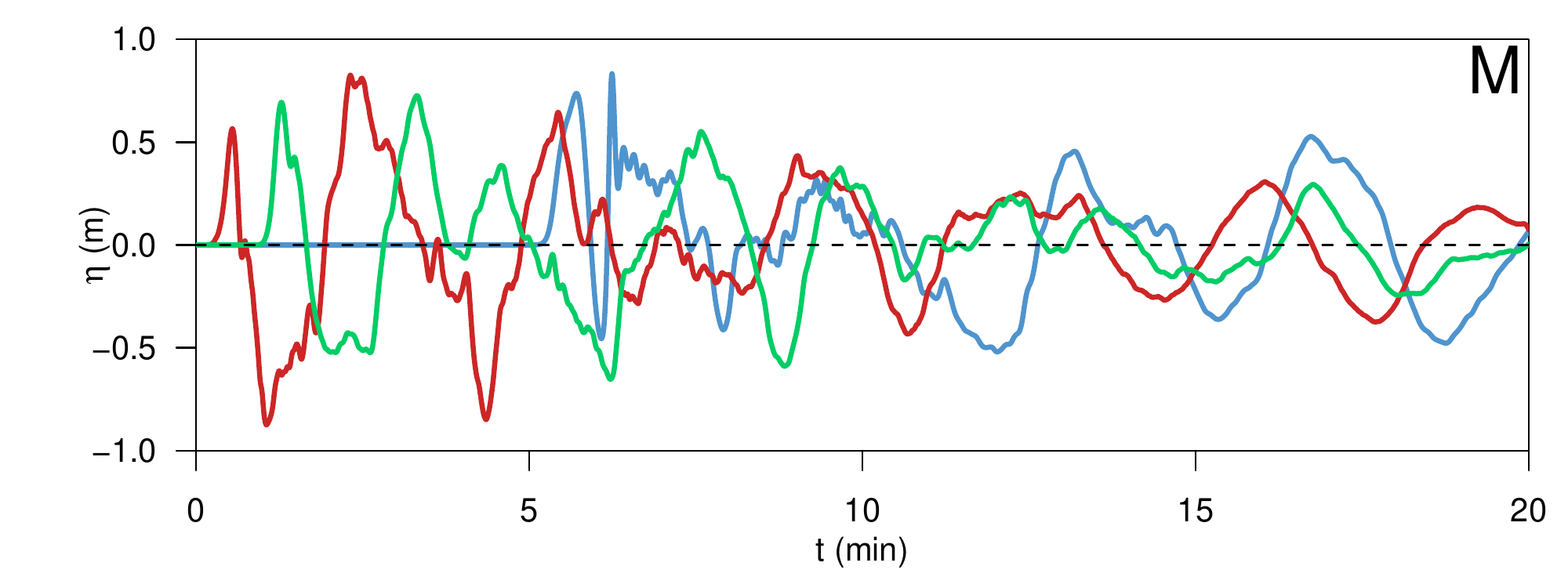}
\includegraphics[width=0.49\textwidth]{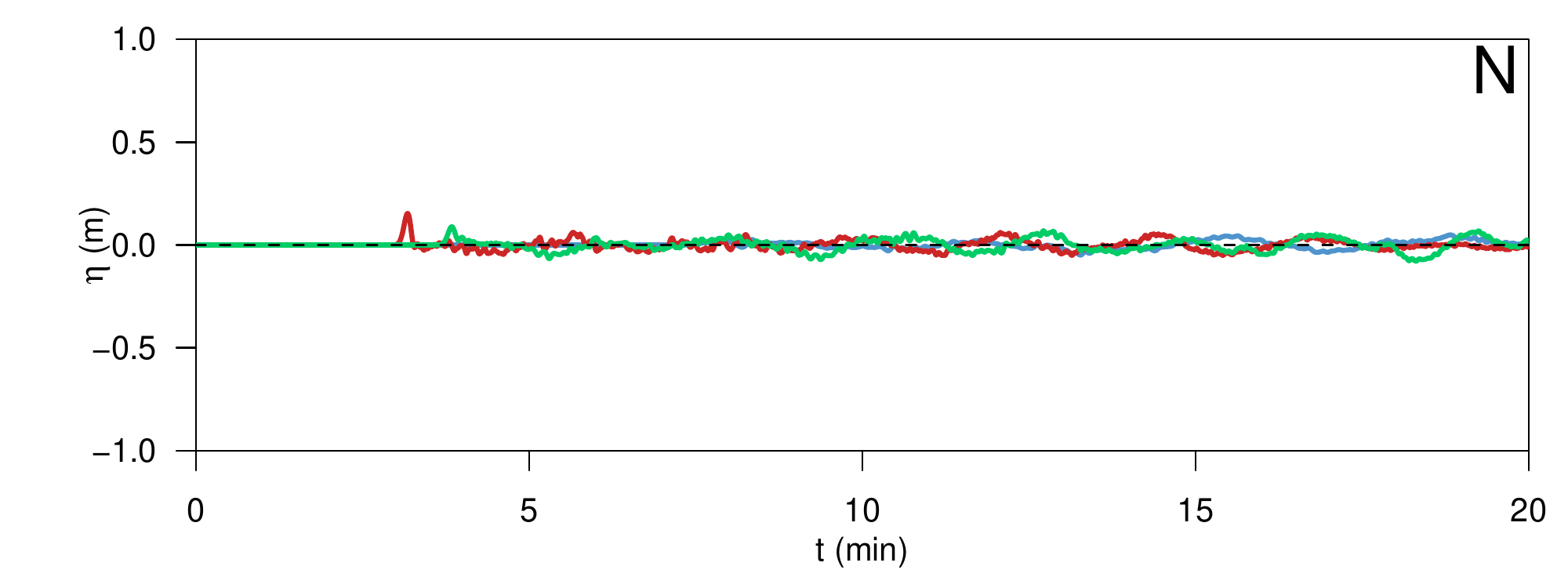}
\caption{Free surface elevation induced independently by each of the landslides at the three locations of interest.}
\label{fig:etaLocations}
\end{figure}

The most relevant results from the simulations are included in Figures~\ref{fig:maxEta} and \ref{fig:etaLocations}.
Figure~\ref{fig:maxEta} shows the highest free surface elevation (FSE) induced across Palu bay and the inland areas inundated over the 20 minutes of the simulation.
Figure~\ref{fig:etaLocations} shows the time series of FSE produced by each of the waves at Palu, Wani and Pantoloan, where data from the tide gauge and \citer{carvajal19} exists.

LGWs from locations A and N barely make any contributions to any of the relevant locations, as expected due to location and orientation and previously reported in \citer{liu20}.
In fact, the LGW A seems to behave like a trapped wave, refracting back to the shoreline instead of propagating towards the deep end of the bay due to the very steep slopes in the area.
LGWs B-F are generated along the west coast of Palu bay, therefore, impact the east coast directly.
The highest wave contribution of B-D at Pantoloand and Wani is between 1 and 2 m, whereas it is much smaller for E and F, which are located further south.
Interestingly, it is very noticeable LGWs B-F are deeply affected by refraction due to the bathymetry, causing the energy of the tsunami to focus at Palu Mall, despite the main wave initially pointing in another direction.
Waves from these landslides arrive at Palu at different times and with maximum amplitudes around 2 m.
LGWs G, H and I occur between Palu's bridge and Talise, with G and H oriented north and I oriented west.
Waves G, H and I arrive at Palu very fast, with maximum positive amplitudes ranging from 0.5 m, 2 m and 6 m, respectively.
The fact that the FSE data reported by \citer{carvajal19} at Palu barely reaches higher than +2 m is an indication that the magnitude of landslide I has been possibly overestimated.
Finally, LGWs J-M are produced at the east coast of the bay, and have diverse initial directions.
LGW J is the closest to Palu and it is oriented in a similar direction, therefore, it induces a wave amplitude of 1 m at that location (energy focussing due to the bathymetry also plays a significant role), and negligible waves at the two other locations.
The contribution of LGWs K, L and M are the largest at Pantoloan and Wani together with B.
Interestingly, LGW L induces a trough-leading wave at Pantoloan very soon after the simulation start, exactly as reported in \citer{carvajal19}, and a very large wave ($~4$ m) at Wani, which may be compatible with a CCTV video of a home recorded 150 m inland \cite{carvajal19}.

Overall, we will consider the FSE results shown in Figure~\ref{fig:etaLocations} as ``orthogonal functions'' to test to what degree can a linear combination of them explain the FSE measurements available.

\subsection{Adjustments in wave characteristics}
\label{sec:adjustments}

An optimisation routine based on the Genetic Algorithm (GA) technique has been programmed to estimate a set of empirical scaling factors to the wave height, wave length and the initiation time of the tsunamis to obtain a closer degree of accordance with the measured/estimated data at Pantoloan and Palu \cite{carvajal19}.
This program minimises a cost function, which is defined as the sum of the square differences between the estimated time series and the data available \cite{carvajal19}.
The new time series are built as a linear combination of the curves shown in Figure~\ref{fig:etaLocations}, where each of the curves is scaled vertically by a factor $f(H)$ and horizontally (in time) by a factor $f(\lambda)$, and shifted in time by a factor $t_s$.
The parameter space for each of these factors has been constrained between 0.25 and 2, between 0.5 and 2 and between 0 and 180 s, respectively.
Following this approach two major assumptions have been made: that wave evolution is linear (e.g., not affected by wave-wave interactions) and that wave period and wavelength are linked linearly, although both assumptions will be relaxed by performing a final simulation with the new estimated parameters.

The cost function has been designed to add together the contributions of Palu and Pantoloan time series, giving a higher weight to Pantoloan's, since the characteristics of the video analysed in \citer{carvajal19} produce smaller uncertainties in the free surface elevation than the videos in Palu.
Moreover, larger weight has also been applied to the first part of the signals, in which the level is 0 because the waves have not arrived yet, to penalise an early arrival of the waves.

The runs have 100 individuals, each with 3 chromosomes and last for a total of 2,000 time steps.
In each step the individuals are sorted from low to high in terms of the cost function.
Then, 25\% of the top individuals are retained, along with 10\% randomly selected from the rest.
Next, 35\% of the total individuals are created by crossing the selected ones (i.e., exchanging chromosomes), with each chromosome having a 20\% chance of mutating randomly.
Finally, the rest of individuals needed to complete the population (30\%) are created randomly.
The whole process takes approximately 4 hours in parallel using 12 cores of a workstation.

\begin{table}
\centering
\caption{Factors of variation for the landslide characteristics to obtain the best fit with the data available.}
\label{tab:landslideChangesGA}
\begin{tabular}{ccccccccccccccc}
\hline\noalign{\smallskip}
 & \textbf{A} & \textbf{B} & \textbf{C} & \textbf{D} & \textbf{E} & \textbf{F} & \textbf{G} & \textbf{H} & \textbf{I} & \textbf{J} & \textbf{K} & \textbf{L} & \textbf{M} & \textbf{N} \\
\noalign{\smallskip}\hline\noalign{\smallskip}
$f(H)$ & 0.33 & 0.40 & 0.81 & 0.82 & 1.19 & 0.47 & 1.39 & 0.61 & 0.32 & 0.48 & 0.71 & 0.51 & 1.28 & 1.37 \\
$f(\lambda)$ & 0.71 & 1.36 & 1.25 & 1.13 & 1.34 & 1.29 & 0.96 & 0.65 & 1.06 & 0.58 & 0.50 & 0.67 & 1.02 & 0.72 \\
$t_{s}$ & 7 & 126 & 33 & 90 & 91 & 177 & 169 & 162 & 165 & 148 & 133 & 140 & 137 & 80 \\
$H_1$ & 1.3 & 2.3 & 4.2 & 8.4 & 3.0 & 6.4 & 8.6 & 4.1 & 2.6 & 0.9 & 2.1 & 5.7 & 5.4 & 1.9 \\
$\lambda_1$ & 85 & 900 & 2025 & 730 & 335 & 660 & 760 & 500 & 885 & 610 & 535 & 925 & 2035 & 530 \\
$t_{s_1}$ & 50 & 75 & 150 & 115 & 10 & 75 & 160 & 155 & 170 & 100 & 150 & 115 & 135 & 50 \\
\noalign{\smallskip}\hline
\end{tabular}
\end{table}
 
The parameters ($f(H)$, $f(\lambda)$ and $t_s$) which yielded the best solution are summarised in Table~\ref{tab:landslideChangesGA}, along with the new wave heights ($H_1$) and wavelengths ($\lambda_1$) to be tested next in new COMCOT simulations.
Overall, the GA optimization technique may indicate that the largest wave heights estimated in Table~\ref{tab:landslideCharacteristics} (D, F, I, L) had been probably overestimated, as was also noted in \citer{liu20}.

\begin{figure}
\centering
\includegraphics[trim=0 36 0 0,clip,width=\textwidth]{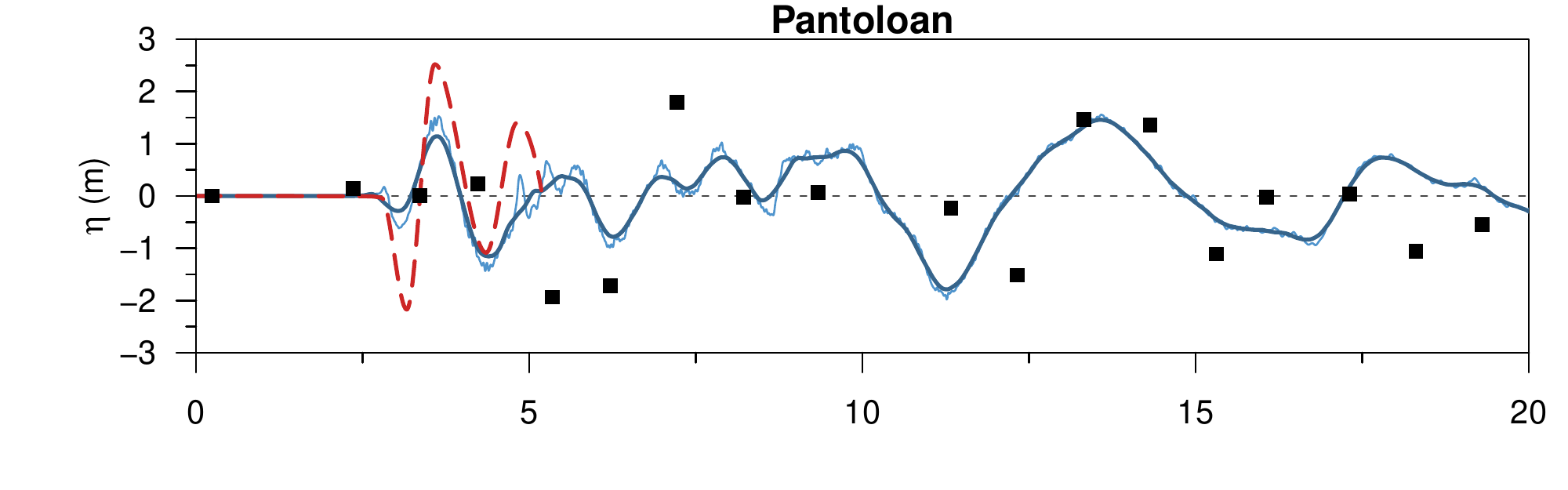}\\
\includegraphics[trim=0 36 0 10,clip,width=\textwidth]{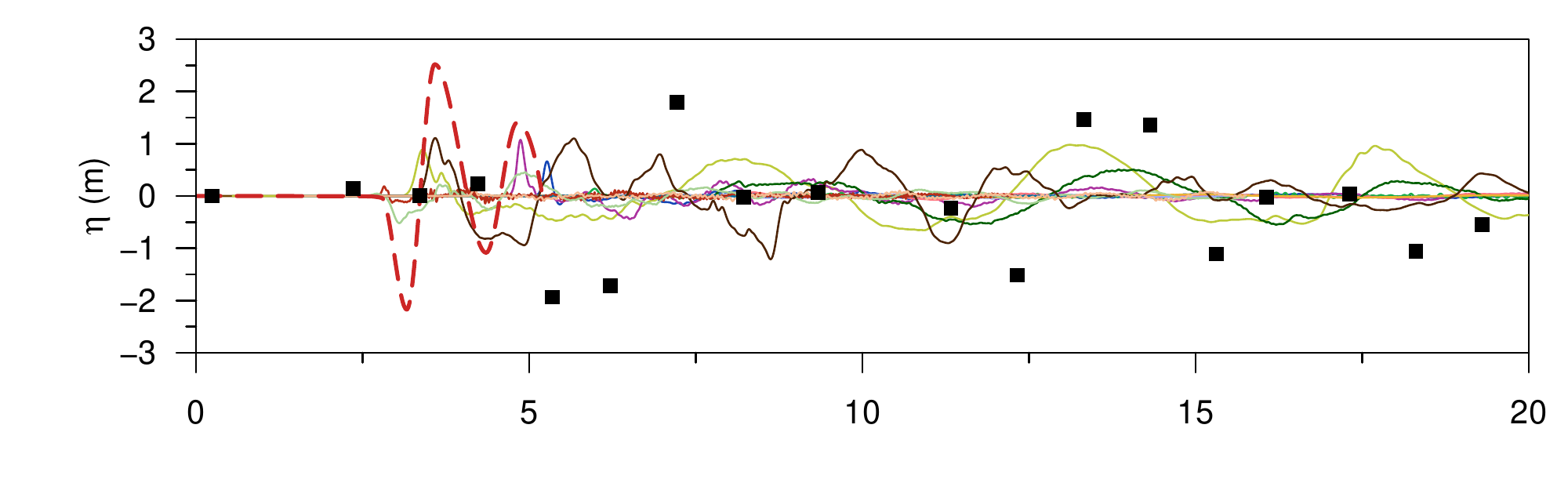}\\
\includegraphics[trim=0 36 0 0,clip,width=\textwidth]{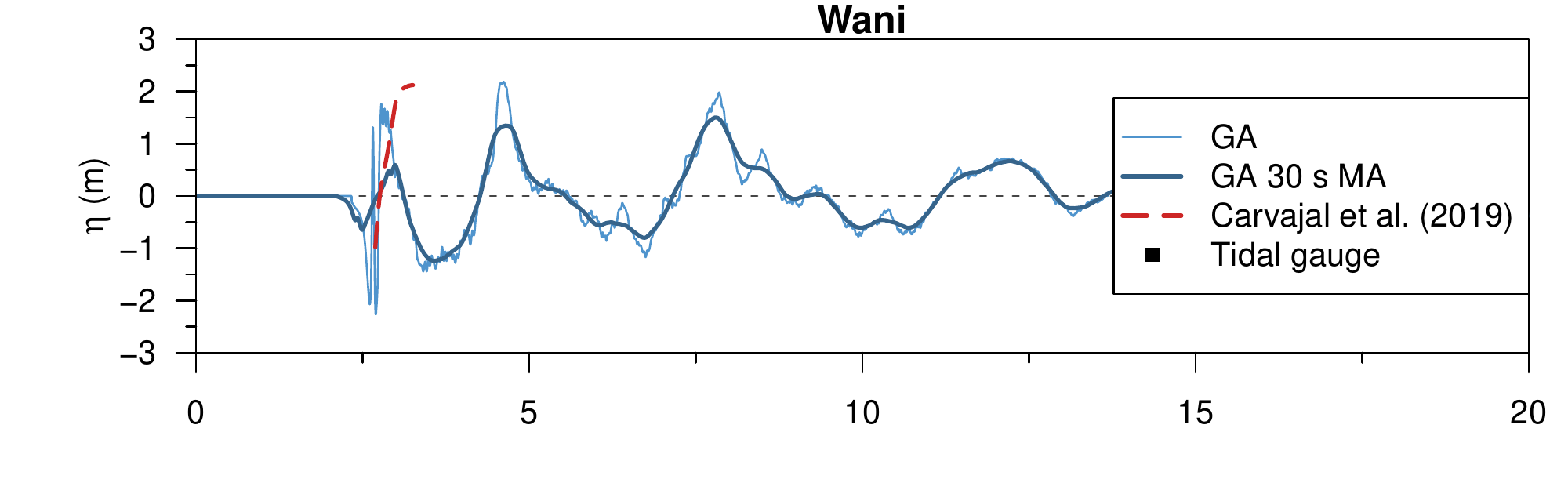}\\
\includegraphics[trim=0 36 0 10,clip,width=\textwidth]{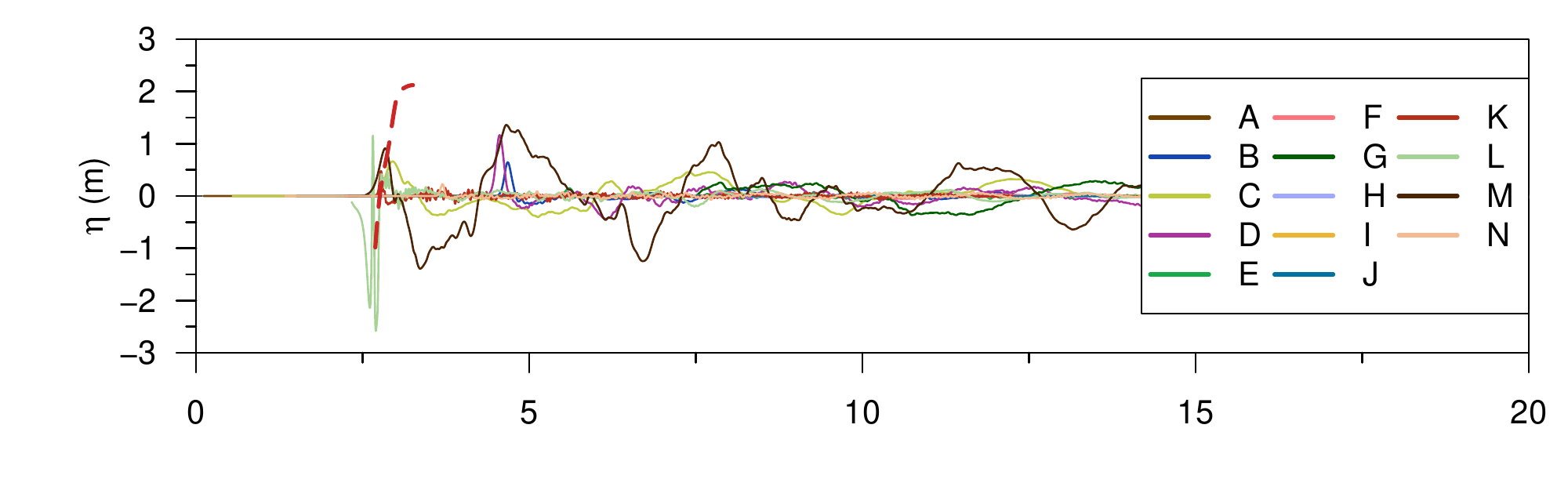}\\
\includegraphics[trim=0 36 0 0,clip,width=\textwidth]{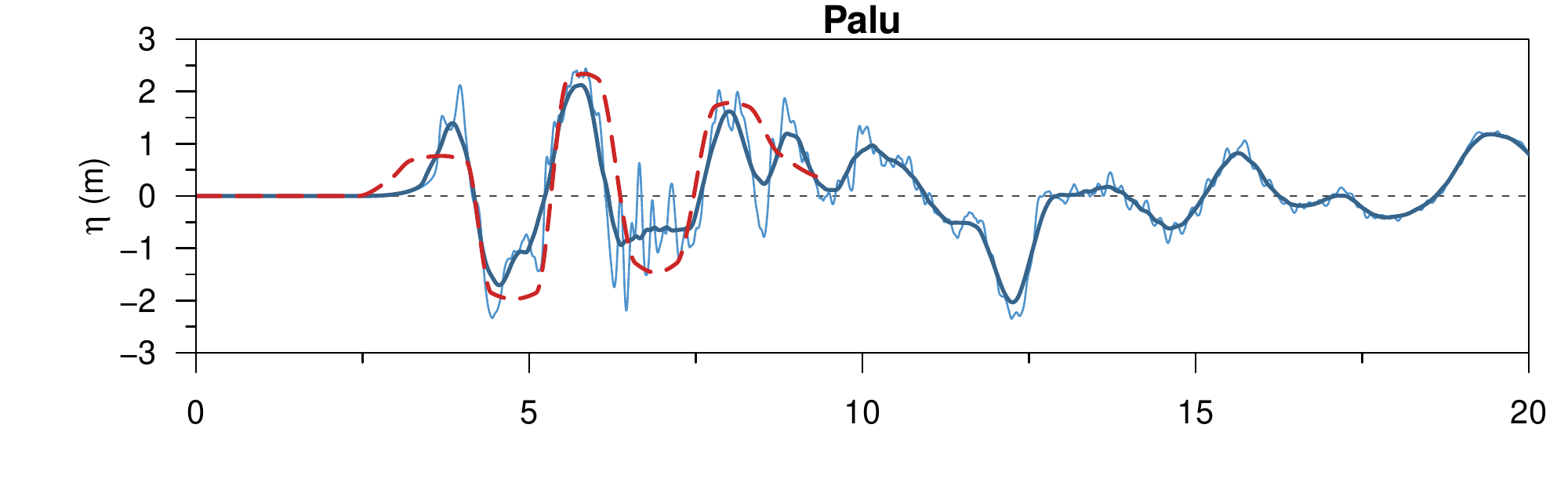}\\
\includegraphics[trim=0 5 0 10,clip,width=\textwidth]{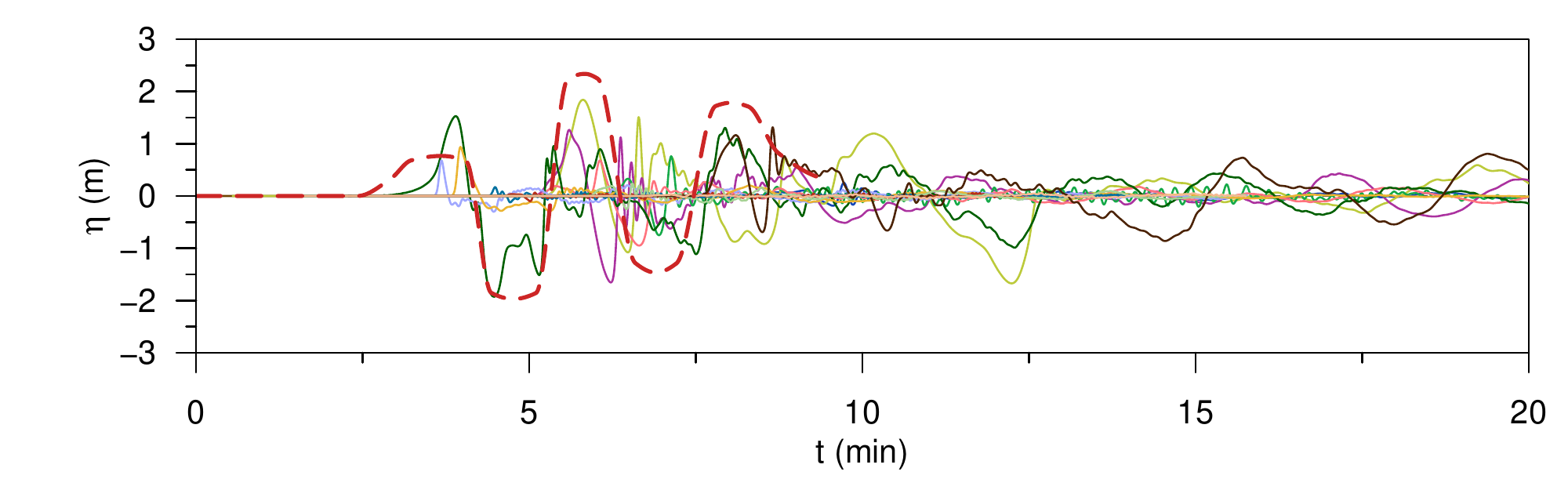}
\caption{Free surface elevation at Pantoloan, Wani and Palu with the best fit calculated by the GA method. Total contribution (odd number panels) and individual landslide contribution (even number panels).}
\label{fig:etaLocationsChangesGA}
\end{figure}

The results of the GA technique can also be represented graphically, as displayed in Figure~\ref{fig:etaLocationsChangesGA}, where the odd number panels show the global result from adding all the contributions from landslides A-N at Pantoloan, Wani and Palu compared with the data presented in \citer{carvajal19}.
The even number panels show the contributions of the individual landslides before aggregating them.
The results in Pantoloan capture well the arrival time of the first wave, which is lead by trough, although the magnitude is much smaller, due to scaling the magnitude of the wave height of landslide L down.
The first wave crest and the following trough are also captured well, although the second crest occurs slightly later.
The time series in Wani is also led by a trough and has a significantly large peak, with magnitude and occurring time very close to those estimated in \citer{carvajal19} from the CCTV of a house.
The greatest contributors to the initial waves are landslides L and M, which are the closest, and C, which is right opposite to Wani.
We note that the estimated time series in Wani is so short that it has not been used in the optimization process, therefore, the matching is reassuring.
Finally, the agreement in Palu is the best among all.
The first wave has a leading crest, originated by the combination of landslides G, H and I, the closest to Palu, with the wave created by G largely contributing to the first wave tough too.
The second crest is mainly produced by the waves from landslides C, D and G (which is now reflected), while the third crest has a primary contribution from landslide M and a subsequent re-reflection of wave the wave from G.

Overall, the optimization technique has shown a satisfactory performance, with promising results to be further evaluated after new COMCOT simulations.
The new wave characteristics presented in Table~\ref{tab:landslideChangesGA} have been simulated individually again to obtain the new set of ``orthogonal functions'', since after changing both wave heights and wavelengths, the FSE at all locations will likely be very different.
With the new data from the 14 simulations, the GA optimization program has been run again, but this time only being able to adjust the lag, in order to preserve the wave height and wavelength calculated before.
This is because the new wave characteristics impact the shape and location of the LGWs, which are calculated using the method of \citer{loPhD}, therefore, even if COMCOT does not account for amplitude dispersion, the arrival times of the waves will be slightly different than before.
The new optimal lags (rounded to the closest 5 seconds for practical reasons) are denoted as $t_{s_1}$ in Table~\ref{tab:landslideChangesGA}.
Finally, all the waves have been reproduced in a single COMCOT simulation, whose goal is to capture the nonlinear effects due to wave-wave interactions that have been disregarded so far because of the initial assumptions.
In this simulation each of the waves is introduced at the time $t_{s_1}$.
Doing so requires stopping and restarting the simulation when adding the new waves, and finally gathering and combining all the data, which is presented in the next section.

\section{Results}
\label{sec:results}

\begin{figure}
\centering
\includegraphics[width=0.91\textwidth]{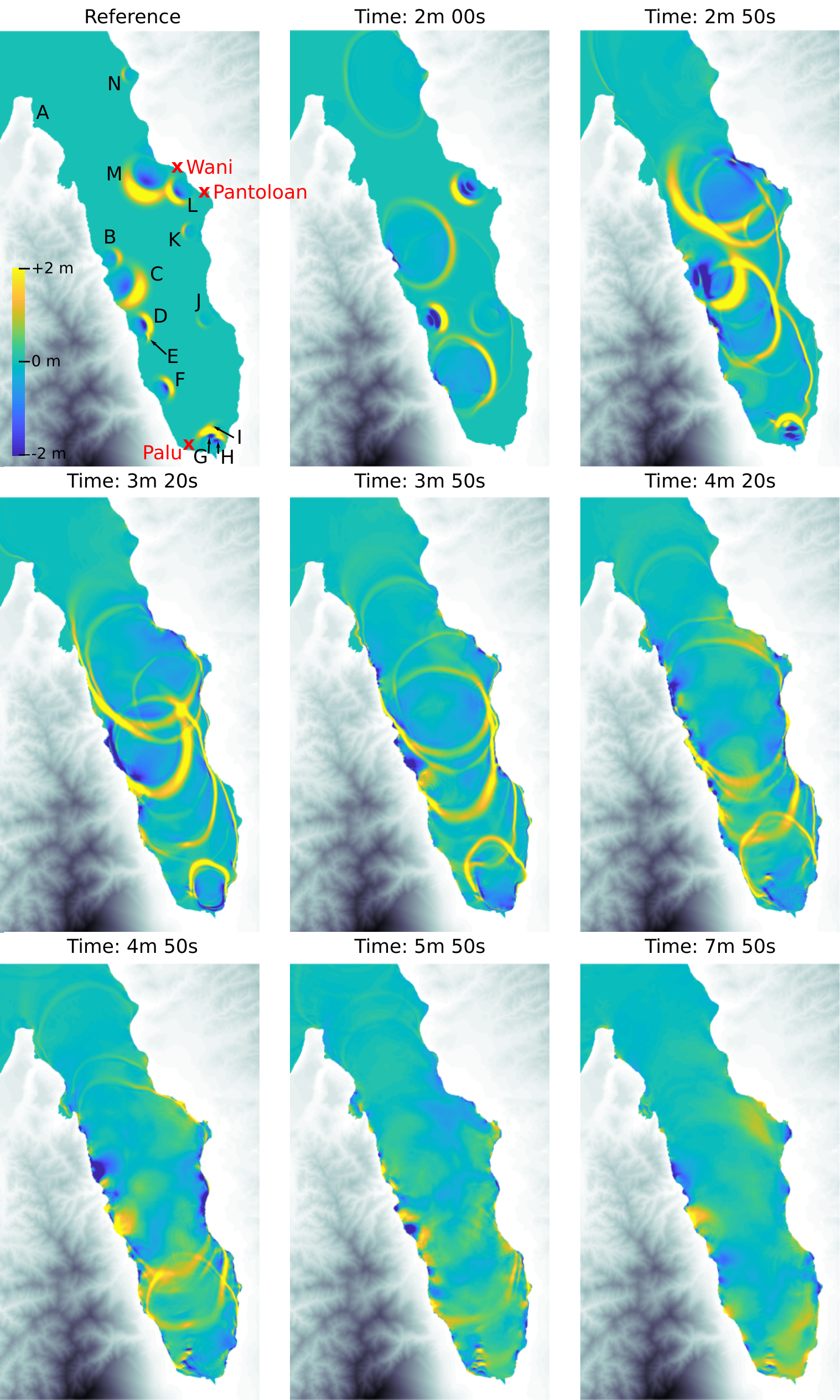}
\caption{Free surface elevation throughout Palu bay at different stages of the simulation.}
\label{fig:tsunamiEvolution}
\end{figure}

\begin{figure}
\centering
\includegraphics[trim=0 35 0 0,clip,width=\textwidth]{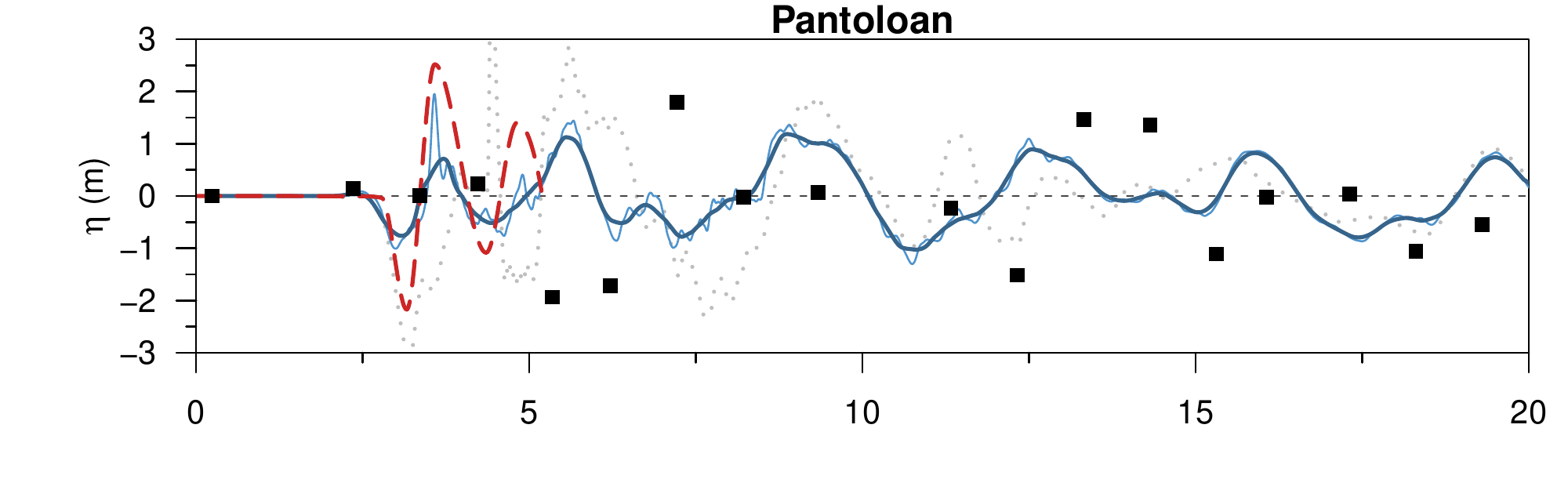}\\
\includegraphics[trim=0 35 0 0,clip,width=\textwidth]{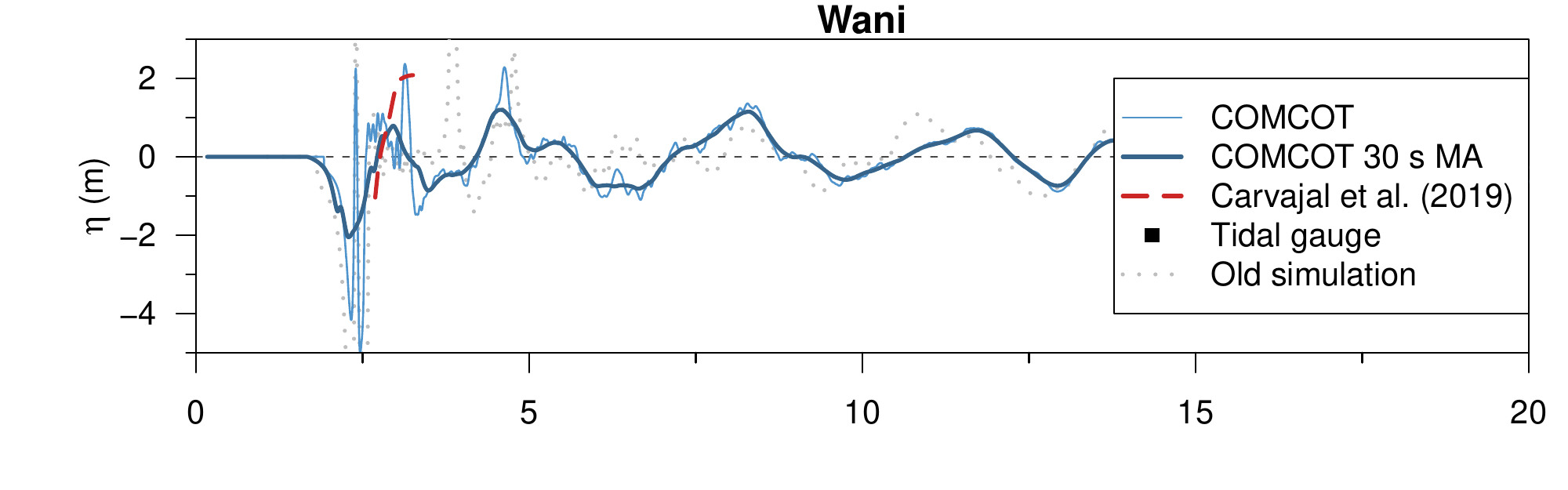}\\
\includegraphics[width=\textwidth]{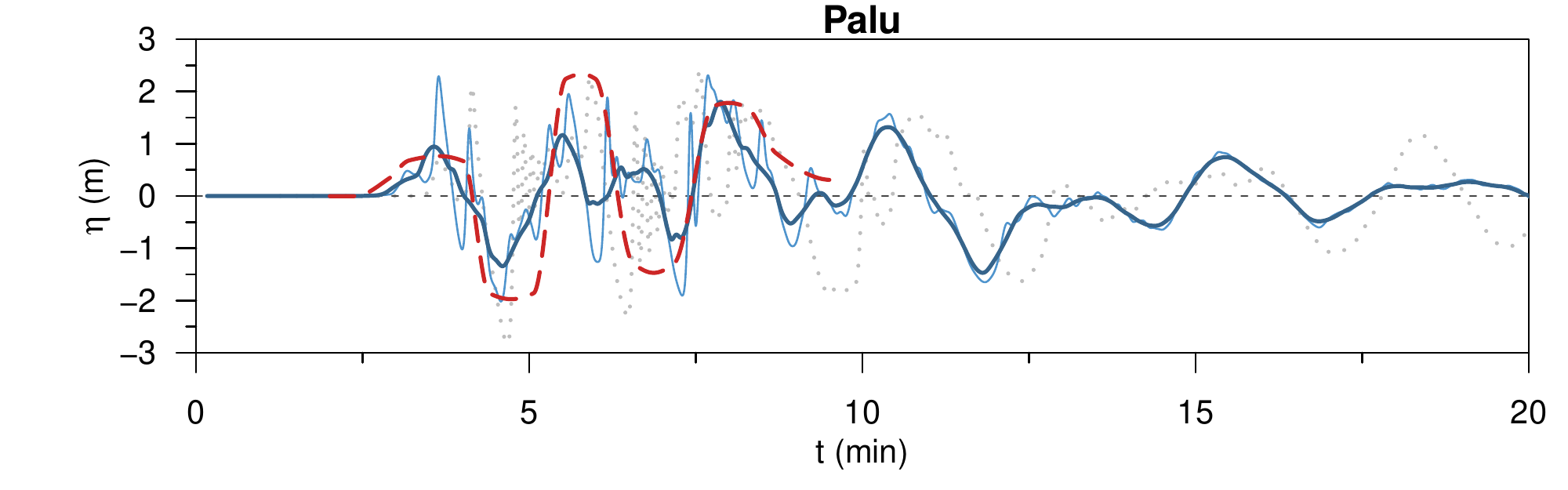}
\caption{Free surface elevation at Pantoloan, Wani and Palu from the final simulation in which all the landslides are included.}
\label{fig:etaLocationsFinal}
\end{figure}

The main results from the final simulation are shown in Figures~\ref{fig:tsunamiEvolution} and \ref{fig:etaLocationsFinal}.

Figure~\ref{fig:tsunamiEvolution} shows the evolution of the LGWs on the bay at different stages during the simulation.
The panel on the top left is for reference only and includes the initial stage of all the LGWs, which are then introduced progressively at the time $t_{s_1}$ in Table~\ref{tab:landslideChangesGA}.
Because of this, the snapshot at time equals 2:00 minutes does not include all the waves yet, since they have just finished being introduced at time 2:50 (top right panel).
By that time the wave N (which is short and small) is propagating out of the bay, waves F and D are about to reach the east coast and the largest waves (C, D, L and M) are still propagating across the bay.
By times 3:20 and 3:50, most of the waves have started propagating in the north to south direction towards Palu, forming long wave fronts and experiencing diffraction due to the bathymetry.
The cluster of waves generated at landslides G, H and I radiates from the south end of the bay, propagating north and impacting the south-most eastern and western shorelines.
However, the most interesting effect is observing those clustered waves bend due to heavy refraction in front of Palu Mall and start propagating south (time 4:20), finally reaching the coast.
The snapshots at times 4:50 and 5:50 show the final waves arriving at Palu, and how they become undular bores as they propagate over the shoal in front of Palu Mall.
Moreover, an oscillatory pattern in the FSE can be observed adjacent to the eastern and western coasts.
These oscillations resemble trapped waves due to the shallow coastal shelf and very short transition to the deep end of the bay (see the steepness of the profile in Figure~\ref{fig:map}).
These trapped waves seem to have between 2.5 and 5 minutes of wave period and at times appear to propagate along the coastal shelf from north to south as edge waves would do.
Such oscillations take a long time to dissipate, as they are still noticeable at time 7:50 and at the end of the 20-minute simulation.

Figure~\ref{fig:etaLocationsFinal} shows the time series of FSE simulated at Pantoloan, Wani and Palu along with the data from \citer{carvajal19}.
The figure also includes the old data from \citer{liu20} in gray dotted line, to evaluate the differences that the changes in the magnitude and timing of the LGWs make.
Moreover, comparisons can be also made with Figure~\ref{fig:etaLocationsChangesGA}, which is the idealised target applying the linearity assumption.

The tsunami waves in Pantoloan start with a trough, which as also shown in Figure~\ref{fig:etaLocationsChangesGA} is not as shallow as the wave measured in place.
The first crest takes place at the correct timing and has a comparable magnitude, although it is slightly shorter compared with the measurements.
The second trough is reasonably well timed, and although the magnitude of the third crest is well captured (improving the estimates in Figure~\ref{fig:etaLocationsChangesGA}), the timing is delayed.
When compared with the old simulation data \cite{liu20}, scaling the waves has resulted in less extreme free surface elevation variations.
At the same time, the oscillations after the first tsunami waves do not match the data from Pantoloan tide gauge, although they show similar long periods ($~4$ minutes), which match the period of the first mode of east-west natural resonance oscillation, as previously noted \cite{liu20}.

The evolution of the tsunami waves in Wani is very similar to the one discussed in Figure~\ref{fig:etaLocationsChangesGA}, and has a compatible event with the inundation shown in CCTV.
Again, the scaling of the waves has resulted in less extreme free surface elevation variations as compared to the old simulation data \cite{liu20}.
Wani and Pantoloan are very close, therefore, the long-period oscillations are very similar, both in terms of amplitude and period.
Moreover, since the two locations are very close to landslides L and M, the simplified approach to introduce the waves once they become independent of the landslide \cite{lo17} may not be the most suitable, as the detailed evolution during the first stages of wave development may be of importance too to reproduce the initial wave arrival.
However, obtaining such details will require a significant increase in computational cost, as previously discussed.

The initial tsunami wave in Palu, caused by landslides G-I, induces an increase in the water level which is well captured in Figure~\ref{fig:etaLocationsFinal} and very close to the measurements \cite{carvajal19}.
However, the evolution in the final simulation is less ideal than the that estimated in Figure~\ref{fig:etaLocationsChangesGA}, indicating that nonlinear effects may be most important in Palu.
This is reasonable, since the local bathymetry is very complex and LGWs become undular bores in the shallow area in front of the mall.
Overall, the second wave crest and both troughs at either side lack the long-wave shape and height of \citer{carvajal19} measurements, but feature oscillations that are not present on the red dashed line but can be seen in the videos recorded at Palu Mall.

\begin{figure}
\centering
\includegraphics[trim=120 40 120 10,clip,width=\textwidth]{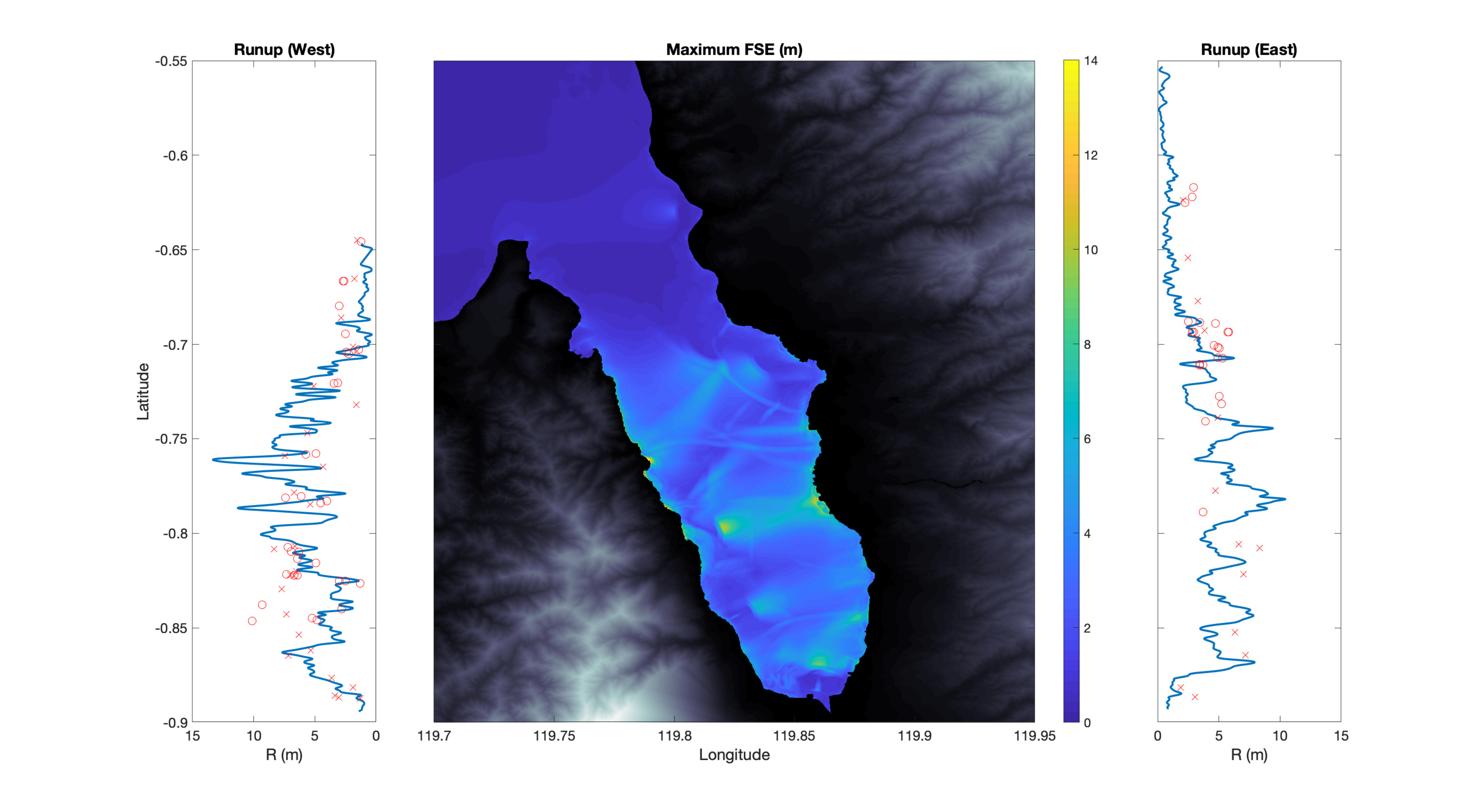}
\caption{Maximum free surface elevation in Palu bay produced by the final simulation (central panel). Analysis of runup at the east and west coasts in the lateral panels. Data from surveys in red markers (circles from \citer{omira19}, crosses from \citer{fritz18}).}
\label{fig:runupFinal}
\end{figure}

Finally, Figure~\ref{fig:runupFinal} shows the maximum free surface elevation throughout Palu bay during the final simulation, along with the runup analysis at the east and west coasts of the bay.
In the central panel, the marks of the initiation stages of the tsunamis are clearly visible, because the waves would start travelling towards deeper waters, thus decreasing in height due to reverse shoaling.
Certain points in which waves crossed one another (e.g., near landslides L and M) are also evident.
However, it is often close to the coast where the amplitudes are the largest, where the shoaling effects are maximal and runup occurs.
The analysis of the simulation yields maximum runups of 13 m in the west coast and 10 m in the east coast, which are only 30\% larger than the maximum values measured at both locations, which do not necessarily occur at the same locations, since runup shows a very localised nature.
This point is supported by the large scatter that the measurements from surveys present.
Overall, the numerically simulated runup shows an adequate agreement with the field measurements, which may strengthen the hypothesis that Palu tsunamis were mainly caused by coastal landslides.

\section{Conclusions and future work}
\label{sec:conclusions}

Due to the limitations in the data available, especially bathymetric survey data before the 2018 event, it will be difficult to get a full understanding and a perfect fit to the measurements of Palu bay tsunami.
In view of the literature review and the numerical modelling data presented in this chapter, the most relevant conclusions that we can draw are as follows.

Video evidence \cite{carvajal19} indicates that the Pantoloan tide gauge did not capture the initial waves arriving at the port, which are likely caused by landslides L and M.
It is often discussed that this may be because the waves were too short to be captured by the low-frequency sampling rate and averaging process.
Moreover, the large amplitude and long period waves measured later at that location are compatible with LGWs exciting the east-west resonant mode of the bay.
The waves measured in Palu can also be explained to a great extent by the landslides, being those the closest to the mall (G, H and I) the most probable contributors to the initial waves that reached the area just 3 minutes after the earthquake struck.
Additionally, the runup data produced by the simulation is coherent with field measurements.
Despite some disagreements in the shape of the runup profile, we believe that agreement may be improved with more detailed simulations of the landslide progression or with a finer tuning of landslide timing \cite{sepulveda20}.
In view of the results the final conclusion would be that coastal landslides can explain most of the measurements and phenomena observed and thus, they should be considered the main source of Palu tsunami.
These results can also be considered a proof against the arguments invalidating the landslide source hypothesis in some references (e.g., \cite{song19, ulrich19}) which often ignore the important role played by the timing of the landslides.

Regarding any other secondary sources, the Pantoloan tide gauge record shows no significant co-seismic deformation, whereas in references sustaining the co-seismic hypothesis modelling usually shows a co-seismic deformation of half a meter at that location.
In the authors' opinion, even if the contribution of the co-seismic deformation cannot be completely ruled out, we can also conclude that this had most probably been secondary and that further developments in this hypothesis may require a completely new, more complex, earthquake model and validation in order to be able to assess its relative importance.
In that sense, although ideally the fault portion located below the sea would need to be surveyed and mapped accurately, the recent work by \citer{natawidjaja20} may be the most promising starting point to gain more insights about the faulting beneath the bay as reflected by the seafloor geometry.
Another hypothesis, submarine landslides deep within the bay \cite{heidarzadeh19} can probably at this point be discarded, since there is no evidence of large enough events in the detailed post-event bathymetry surveys \cite{frederik19, liu20}.

Finally we would like to conclude this chapter by outlining future work that will help improve the understanding of Palu tsunami event.
The most relevant item, which has already been discussed, is performing a detailed mapping of the underwater portion of the fault, which will help to quantify the contribution of the co-seismic deformation hypothesis.
Additionally, tsunami models including sophisticated governing equations have demonstrated to be relevant to accurately model the Palu tsunami \cite{schambach21}, as they can introduce frequency and amplitude dispersion.
In that sense employing a Boussinesq wave-based model in the future instead of the nonlinear shallow water wave solver COMCOT will be advantageous and a step forward to fully explore the potential of the methodology presented.

\subsection*{Acknowledgements}

P. L.-F. Liu would like to acknowledge the funding from the National University of Singapore to support this research.

\backmatter
\bibliographystyle{plainnat_springer}
\bibliography{bibliographyBook}

\begin{thebibliography}{40}
\providecommand{\natexlab}[1]{#1}
\providecommand{\url}[1]{\texttt{#1}}
\expandafter\ifx\csname urlstyle\endcsname\relax
  \providecommand{\doi}[1]{doi: #1}\else
  \providecommand{\doi}{doi: \begingroup \urlstyle{rm}\Url}\fi

\bibitem[Ar\'{a}nguiz et~al.(2020)Ar\'{a}nguiz, Esteban, Takagi, Mikami,
  Takabatake, G\'{o}mez, Gonz\'{a}lez, Shibayama, Okuwaki, Yagi, Shimizu,
  Achiari, Stolle, Robertson, Ohira, Nakamura, Nishida, Krautwald, Goseberg,
  and Nistor]{aranguiz20}
R.~Ar\'{a}nguiz, M.~Esteban, H.~Takagi, T.~Mikami, T.~Takabatake, M.~G\'{o}mez,
  J.~Gonz\'{a}lez, T.~Shibayama, R.~Okuwaki, Y.~Yagi, K.~Shimizu, H.~Achiari,
  J.~Stolle, I.~Robertson, K.~Ohira, R.~Nakamura, Y.~Nishida, C.~Krautwald,
  N.~Goseberg, and I.~Nistor, (2020).
\newblock The 2018 {Sulawesi tsunami in Palu} city as a result of several
  landslides and coseismic tsunamis.
\newblock \emph{Coastal Engineering Journal}, 62\penalty0 (4):\penalty0
  445--459.
\newblock
  \href{https://doi.org/10.1080/21664250.2020.1780719}{https://doi.org/10.1080/21664250.2020.1780719}.

\bibitem[Arikawa et~al.(2018)Arikawa, Muhari, Okumura, Dohi, Afriyanto,
  Sujatmiko, and Imamura]{arikawa18}
T.~Arikawa, A.~Muhari, Y.~Okumura, Y.~Dohi, B.~Afriyanto, K.~A. Sujatmiko, and
  F.~Imamura, (2018).
\newblock Coastal subsidence induced several tsunamis during the 2018
  {Sulawesi} earthquake.
\newblock \emph{Journal of Disaster Research}, 13:\penalty0 sc20181204.
\newblock
  \href{https://doi.org/10.20965/jdr.2018}{https://doi.org/10.20965/jdr.2018}.

\bibitem[Bao et~al.(2019)Bao, Ampuero, Meng, Fielding, Liang, Milliner, Feng,
  and Huang]{bao19}
H.~Bao, J.‐P. Ampuero, L.~Meng, E.~J. Fielding, C.~Liang, C.~W.~D. Milliner,
  T.~R. Feng, and H.~Huang, (2019).
\newblock Early and persistent supershear rupture of the 2018 magnitude 7.5
  {Palu} earthquake.
\newblock \emph{Nature Geoscience}, 12:\penalty0 200--205.
\newblock
  \href{https://doi.org/10.1038/s41561-018-0297-z}{https://doi.org/10.1038/s41561-018-0297-z}.

\bibitem[Bellier et~al.(2001)Bellier, S\'{e}brier, Beaudouin, Villeneuve,
  Braucher, Bourles, Siame, Putranto, and Pratomo]{bellier01}
O.~Bellier, M.~S\'{e}brier, T.~Beaudouin, M.~Villeneuve, R.~Braucher,
  D.~Bourles, L.~Siame, E.~Putranto, and I.~Pratomo, (2001).
\newblock High slip rate for a low seismicity along the {Palu‐Koro} active
  fault in central {Sulawesi (Indonesia)}.
\newblock \emph{Terra Nova}, 13:\penalty0 463--470.
\newblock
  \href{https://doi.org/10.1046/j.1365-3121.2001.00382.x}{https://doi.org/10.1046/j.1365-3121.2001.00382.x}.

\bibitem[Carvajal et~al.(2019)Carvajal, Araya-Cornejo, Sep\'{u}lveda, Melnick,
  and Haase]{carvajal19}
M.~Carvajal, C.~Araya-Cornejo, I.~Sep\'{u}lveda, D.~Melnick, and J.~S. Haase,
  (2019).
\newblock Nearly-instantaneous tsunamis following the {Mw 7.5 2018 Palu}
  earthquake.
\newblock \emph{Geophysical Research Letters}, 46:\penalty0 5117--5126.
\newblock
  \href{https://doi.org/10.1029/2019GL082578}{https://doi.org/10.1029/2019GL082578}.

\bibitem[Frederik et~al.(2019)Frederik, Udrekh, Adhitama, Hananto, Asrafil,
  Sahabuddin, Irfan, Moefti, Putra, and Riyalda]{frederik19}
M.~C.~G. Frederik, Udrekh, R.~Adhitama, N.~D. Hananto, Asrafil, S.~Sahabuddin,
  M.~Irfan, O.~Moefti, D.~B. Putra, and B.~F. Riyalda, (2019).
\newblock First results of a bathymetric survey of {Palu Bay, Central Sulawesi,
  Indonesia} following the tsunamigenic earthquake of 28 {September} 2018.
\newblock \emph{Pure and Applied Geophysics}, 176:\penalty0 3277--3290.
\newblock
  \href{https://doi.org/10.1007/s00024-019-02280-7}{https://doi.org/10.1007/s00024-019-02280-7}.

\bibitem[Fritz et~al.(2013)Fritz, Hillaire, Moli\`{E}re, Wei, and
  Mohammed]{fritz13}
H.~M. Fritz, J.~V. Hillaire, E.~Moli\`{E}re, Y.~Wei, and F.~Mohammed, (2013).
\newblock Twin tsunamis triggered by the 12 {January 2010 Haiti} earthquake.
\newblock \emph{Pure and Applied Geophysics}, 170:\penalty0 1463--1474.
\newblock
  \href{https://doi.org/10.1007/s00024-012-0479-3}{https://doi.org/10.1007/s00024-012-0479-3}.

\bibitem[Fritz et~al.(2018)Fritz, Synolakis, Kalligeris, Skanavis,
  Mohammad~Rizal, Prasetya, and Liu]{fritz18}
H.~M. Fritz, C.~Synolakis, N.~Kalligeris, V.~Skanavis, F.~S. Mohammad~Rizal,
  G.~S. Prasetya, and P.~L.-F. Liu, (2018).
\newblock Field survey of the 28 {September} 2018 {Sulawesi} tsunami.
\newblock In \emph{American Geophysical Union Fall Meeting 2018}.
\newblock Abstract NH22B-04.

\bibitem[Garzon and Ferreira(2016)]{garzon16}
J.~Garzon and C.~Ferreira, (2016).
\newblock Storm surge modeling in large estuaries: sensitivity analyses to
  parameters and physical processes in the {Chesapeake Bay}.
\newblock \emph{Journal of Marine Science and Engineering}, 4.
\newblock
  \href{https://doi.org/10.3390/jmse4030045}{https://doi.org/10.3390/jmse4030045}.

\bibitem[Ghaïtanellis et~al.(2021)Ghaïtanellis, Violeau, Liu, and
  Viard]{ghaitanellis21}
A.~Ghaïtanellis, D.~Violeau, P.~L.‐F. Liu, and T.~Viard, (2021).
\newblock Sph simulation of the 2007 chehalis lake landslide and subsequent
  tsunami.
\newblock \emph{Journal of Hydraulic Research}, In Press.
\newblock
  \href{https://doi.org/10.1080/00221686.2020.1844814}{https://doi.org/10.1080/00221686.2020.1844814}.

\bibitem[Harbitz et~al.(2006)Harbitz, Løvholt, Pedersen, and
  Masson]{harbitz06}
C.~B. Harbitz, F.~Løvholt, G.~Pedersen, and D.~G. Masson, (2006).
\newblock Mechanisms of tsunami generation by submarine landslides: a short
  review.
\newblock \emph{Norwegian Journal of Geology}, 86:\penalty0 255--264.

\bibitem[Heidarzadeh et~al.(2019)Heidarzadeh, Muhari, and
  Wijanarto]{heidarzadeh19}
M.~Heidarzadeh, A.~Muhari, and A.~B. Wijanarto, (2019).
\newblock Insights on the source of the 28 {September} 2018 {Sulawesi} tsunami,
  {Indonesia}, based on spectral analyses and numerical simulations.
\newblock \emph{Pure and Applied Geophysics}, 176:\penalty0 25--42.
\newblock
  \href{https://doi.org/10.1007/s00024-018-2065-9}{https://doi.org/10.1007/s00024-018-2065-9}.

\bibitem[Jaya et~al.(2019)Jaya, Nishikawa, and Jumadil]{jaya19}
A.~Jaya, O.~Nishikawa, and S.~Jumadil, (2019).
\newblock Distribution and morphology of the surface ruptures of the 2018
  {Donggala–Palu earthquake, Central Sulawesi, Indonesia.}
\newblock \emph{Earth Planets Space}, 71.
\newblock
  \href{https://doi.org/10.1186/s40623-019-1126-3}{https://doi.org/10.1186/s40623-019-1126-3}.

\bibitem[Kamphuis and Bowering(1970)]{kamphuis70}
J.~W. Kamphuis and R.~J. Bowering, (1970).
\newblock Impulse waves generated by landslides.
\newblock In \emph{Proceedings of the 12th International Conference on Coastal
  Engineering, Washington D.C., USA}, pages 575--588.

\bibitem[Liu et~al.(2020)Liu, Higuera, Husrin, Prasetya, Prihantono, Diastomo,
  Pryambodo, and Susmoro]{liu20}
P.~L.-F. Liu, P.~Higuera, S.~Husrin, G.~S. Prasetya, J.~Prihantono,
  H.~Diastomo, D.~G. Pryambodo, and H.~Susmoro, (2020).
\newblock Coastal landslides in {Palu Bay during 2018 Sulawesi} earthquake and
  tsunami.
\newblock \emph{Landslides}, 17:\penalty0 2085--2098.
\newblock
  \href{https://doi.org/10.1007/s10346-020-01417-3}{https://doi.org/10.1007/s10346-020-01417-3}.

\bibitem[Lo(2018)]{loPhD}
H.~Y. Lo, (2018).
\newblock \emph{Modeling landslide-generated tsunamis with long-wave
  equations.}
\newblock PhD thesis, Cornell University.

\bibitem[Lo and Liu(2017)]{lo17}
H.~Y. Lo and P.~L.-F. Liu, (2017).
\newblock On the analytical solutions for water waves generated by a prescribed
  landslide.
\newblock \emph{Journal of Fluid Mechanics}, 821:\penalty0 85--116.
\newblock
  \href{https://doi.org/10.1017/jfm.2017.251}{https://doi.org/10.1017/jfm.2017.251}.

\bibitem[Mai(2019)]{mai19}
P.~M. Mai, (2019).
\newblock Supershear tsunami disaster.
\newblock \emph{Nature Geoscience}, 12:\penalty0 150--151.
\newblock
  \href{https://doi.org/10.1038/s41561-019-0308-8}{https://doi.org/10.1038/s41561-019-0308-8}.

\bibitem[Mitchell(1998)]{mitchell98}
M.~Mitchell, (1998).
\newblock \emph{An introduction to genetic algorithms.}
\newblock MIT press.

\bibitem[Natawidjaja et~al.(2020)Natawidjaja, Daryono, Prasetya, Udrekh, Liu,
  Hananto, Kongko, Triyoso, Puji, Meilano, Gunawan, Supendi, Pamumpuni, Irsyam,
  Faizal, Hidayati, Sapiie, Kusuma, and Tawil]{natawidjaja20}
D.~H. Natawidjaja, M.~R. Daryono, G.~Prasetya, Udrekh, P.~L.-F. Liu, N.~D.
  Hananto, W.~Kongko, W.~Triyoso, A.~R. Puji, I.~Meilano, E.~Gunawan,
  P.~Supendi, A.~Pamumpuni, M.~Irsyam, L.~Faizal, S.~Hidayati, B.~Sapiie, M.~A.
  Kusuma, and S.~Tawil, (2020).
\newblock {The 2018 Mw7.5 Palu ‘supershear’ earthquake ruptures geological
  fault's multisegment separated by large bends: results from integrating field
  measurements, LiDAR, swath bathymetry and seismic-reflection data.}
\newblock \emph{Geophysical Journal International}, 224\penalty0 (2):\penalty0
  985--1002.
\newblock
  \href{https://doi.org/10.1093/gji/ggaa498}{https://doi.org/10.1093/gji/ggaa498}.

\bibitem[Okada(1985)]{okada85}
Y.~Okada, (1985).
\newblock Surface deformation due to shear and tensile faults in a half-space.
\newblock \emph{Bulletin of the seismological society of America}, 75:\penalty0
  1135--1154.

\bibitem[Omira et~al.(2019)Omira, Dogan, Hidayat, Husrin, Prasetya, Annunziato,
  Proietti, Probst, Paparo, Wronna, Zaytsev, Pronin, Giniyatullin, Putra,
  Hartanto, Ginanjar, Kongko, Pelinovsky, and Yalciner]{omira19}
R.~Omira, G.~G. Dogan, R.~Hidayat, S.~Husrin, G.~S. Prasetya, A.~Annunziato,
  C.~Proietti, P.~Probst, M.~A. Paparo, M.~Wronna, A.~Zaytsev, P.~Pronin,
  A.~Giniyatullin, P.~S. Putra, D.~Hartanto, G.~Ginanjar, W.~Kongko,
  E.~Pelinovsky, and A.~C. Yalciner, (2019).
\newblock The {September} 28th, 2018, tsunami in {Palu-Sulawesi, Indonesia: A}
  post-event field survey.
\newblock \emph{Pure and Applied Geophysics}, 176:\penalty0 1379--1395.
\newblock
  \href{https://doi.org/10.1007/s00024-019-02145-z}{https://doi.org/10.1007/s00024-019-02145-z}.

\bibitem[Pakoksung et~al.(2019)Pakoksung, Suppasri, Imamura, Athanasius, Omang,
  and Muhari]{pakoksung19}
K.~Pakoksung, A.~Suppasri, F.~Imamura, C.~Athanasius, A.~Omang, and A.~Muhari,
  (2019).
\newblock Simulation of the submarine landslide tsunami on 28 {September} 2018
  in {Palu Bay, Sulawesi Island, Indonesia}, using a two-layer model.
\newblock \emph{Pure and Applied Geophysics}, 176:\penalty0 3323--3350.
\newblock
  \href{https://doi.org/10.1007/s00024-019-02235-y}{https://doi.org/10.1007/s00024-019-02235-y}.

\bibitem[Panizzo et~al.(2005)Panizzo, De~Girolamo, and Petaccia]{panizzo05}
A.~Panizzo, P.~De~Girolamo, and A.~Petaccia, (2005).
\newblock Forecasting impulse waves generated by subaerial landslides.
\newblock \emph{Journal of Geophysical Research: Oceans}, 110\penalty0 (C12).
\newblock
  \href{https://doi.org/10.1029/2004JC002778}{https://doi.org/10.1029/2004JC002778}.

\bibitem[Paulik et~al.(2019)Paulik, Gusman, Williams, Pratama, Lin,
  Prawirabhakti, Sulendraand, Zachari, Fortuna, Layuk, and Suwarni]{paulik19}
R.~Paulik, A.~Gusman, J.~H. Williams, G.~M. Pratama, S.-L. Lin,
  A.~Prawirabhakti, K.~Sulendraand, M.~Y. Zachari, Z.~E.~D. Fortuna, N.~B.~P.
  Layuk, and N.~W.~I. Suwarni, (2019).
\newblock Tsunami hazard and built environment damage observations from {Palu
  City after the September 28 2018 Sulawesi} earthquake and tsunami.
\newblock \emph{Pure and Applied Geophysics}, 176:\penalty0 3305--3321.
\newblock
  \href{https://doi.org/10.1007/s00024-019-02254-9}{https://doi.org/10.1007/s00024-019-02254-9}.

\bibitem[Prasetya et~al.(2001)Prasetya, De~Lange, and Healy]{prasetya01}
G.~Prasetya, W.~De~Lange, and T.~Healy, (2001).
\newblock The {Makassar} strait tsunamigenic region, {Indonesia}.
\newblock \emph{Natural Hazards}, 24:\penalty0 295--307.
\newblock
  \href{https://doi.org/10.1023/A:1012297413280}{https://doi.org/10.1023/A:1012297413280}.

\bibitem[Romano(2020)]{romamo20}
A.~Romano, (2020).
\newblock \emph{Physical and Numerical Modeling of Landslide-Generated
  Tsunamis: A Review.}
\newblock
  \href{https://doi.org/10.5772/intechopen.93878}{https://doi.org/10.5772/intechopen.93878}.

\bibitem[Sassa and Takagawa(2019)]{sassa19}
S.~Sassa and T.~Takagawa, (2019).
\newblock Liquefied gravity flow-induced tsunami: first evidence and comparison
  from the 2018 {Indonesia Sulawesi} earthquake and tsunami disasters.
\newblock \emph{Landslides}, 16:\penalty0 195--200.
\newblock
  \href{https://doi.org/10.1007/s10346-018-1114-x}{https://doi.org/10.1007/s10346-018-1114-x}.

\bibitem[Schambach et~al.(2021)Schambach, Grilli, and Tappin]{schambach21}
L.~Schambach, S.~T. Grilli, and D.~R. Tappin, (2021).
\newblock New high-resolution modeling of the 2018 {Palu} tsunami, based on
  supershear earthquake mechanisms and mapped coastal landslides, supports a
  dual source.
\newblock \emph{Frontiers in Earth Science}, 8:\penalty0 627.
\newblock
  \href{https://doi.org/10.3389/feart.2020.598839}{https://doi.org/10.3389/feart.2020.598839}.

\bibitem[Sep\'{u}lveda et~al.(2018)Sep\'{u}lveda, Haase, Liu, Xu, and
  Carvajal]{sepulveda18}
I.~Sep\'{u}lveda, J.~S. Haase, P.~L.-F. Liu, X.~Xu, and M.~Carvajal, (2018).
\newblock On the contribution of co-seismic displacements to the 2018 {Palu}
  tsunami.
\newblock In \emph{American Geophysical Union Fall Meeting 2018}.
\newblock Abstract NH23F-3552.

\bibitem[Sep\'{u}lveda et~al.(2020)Sep\'{u}lveda, Haase, Carvajal, Xu, and
  Liu]{sepulveda20}
I.~Sep\'{u}lveda, J.~S. Haase, M.~Carvajal, X.~Xu, and P.~L.-F. Liu, (2020).
\newblock Modeling the sources of the 2018 {Palu, Indonesia}, tsunami using
  videos from social media.
\newblock \emph{Journal of Geophysical Research: Solid Earth}, 125,
  e2019JB018675.
\newblock
  \href{https://doi.org/10.1029/2019JB018675}{https://doi.org/10.1029/2019JB018675}.

\bibitem[Socquet et~al.(2019)Socquet, Hollingsworth, Pathier, and
  Bouchon]{soquet19}
A.~Socquet, J.~Hollingsworth, E.~Pathier, and M.~Bouchon, (2019).
\newblock Evidence of supershear during the 2018 magnitude 7.5 {Palu}
  earthquake from space geodesy.
\newblock \emph{Nature Geoscience}, 12:\penalty0 192--199.
\newblock
  \href{https://doi.org/10.1038/s41561-018-0296-0}{https://doi.org/10.1038/s41561-018-0296-0}.

\bibitem[Song et~al.(2019)Song, Zhang, Shan, Liu, Gong, and Qu]{song19}
X.~Song, Y.~Zhang, X.~Shan, Y.~Liu, W.~Gong, and C.~Qu, (2019).
\newblock Geodetic observations of the 2018 {Mw 7.5 Sulawesi} earthquake and
  its implications for the kinematics of the {Palu} fault.
\newblock \emph{Geophysical Research Letters}, 46:\penalty0 4212--4220.
\newblock
  \href{https://doi.org/10.1029/2019GL082045}{https://doi.org/10.1029/2019GL082045}.

\bibitem[Song et~al.(2018)Song, Chen, Liu, Roback, and Avouac]{song18}
Y.~T. Song, K.~Chen, Z.~Liu, K.~Roback, and J.~P. Avouac, (2018).
\newblock The 28 {September 2018 Mw 7.5 Sulawesi (Indonesia)} earthquake and
  its implication for tsunami early warning.
\newblock In \emph{American Geophysical Union Fall Meeting 2018}.
\newblock Abstract NH23F-3545.

\bibitem[Synolakis et~al.(2002)Synolakis, Bardet, Borrero, Davies, Okal,
  Silver, Sweet, and Tappin]{synolakis02}
C.~E. Synolakis, J.-P. Bardet, J.~C. Borrero, H.~L. Davies, E.~A. Okal, E.~A.
  Silver, S.~Sweet, and D.~R. Tappin, (2002).
\newblock The slump origin of the 1998 papua new guinea tsunami.
\newblock \emph{Proceedings: Mathematical, Physical and Engineering Sciences},
  458:\penalty0 763--789.
\newblock
  \href{https://doi.org/10.1098/rspa.2001.0915}{https://doi.org/10.1098/rspa.2001.0915}.

\bibitem[Takagi et~al.(2019)Takagi, Pratama, Kurobe, Esteban, Aránguiz, and
  Ke]{takagi19}
H.~Takagi, M.~B. Pratama, S.~Kurobe, M.~Esteban, R.~Aránguiz, and B.~Ke,
  (2019).
\newblock Analysis of generation and arrival time of landslide tsunami to {Palu
  City} due to the 2018 {Sulawesi} earthquake.
\newblock \emph{Landslides}, 16:\penalty0 983--991.
\newblock
  \href{https://doi.org/10.1007/s10346-019-01166-y}{https://doi.org/10.1007/s10346-019-01166-y}.

\bibitem[Ulrich et~al.(2019)Ulrich, Vater, Madden, Behrens, van Dinther, van
  Zelst, Fielding, Liang, and Gabriel]{ulrich19}
T.~Ulrich, S.~Vater, E.~H. Madden, J.~Behrens, Y.~van Dinther, I.~van Zelst,
  E.~J. Fielding, C.~Liang, and A.-A. Gabriel, (2019).
\newblock Coupled, physics-based modeling reveals earthquake displacements are
  critical to the 2018 {Palu, Sulawesi} tsunami.
\newblock \emph{Pure and Applied Geophysics}, 176:\penalty0 4069--4109.
\newblock
  \href{https://doi.org/10.1007/s00024-019-02290-5}{https://doi.org/10.1007/s00024-019-02290-5}.

\bibitem[Valkaniotis et~al.(2018)Valkaniotis, Ganas, Tsironi, and
  Barberopoulou]{valkaniotis18}
S.~Valkaniotis, A.~Ganas, V.~Tsironi, and A.~Barberopoulou, (2018).
\newblock A preliminary report on the {M7.5 Palu} earthquake co-seismic
  ruptures and landslides using image correlation techniques on optical
  satellite data.
\newblock \emph{Technical report}.
\newblock
  \href{http://dx.doi.org/10.5281/zenodo.1467128}{http://dx.doi.org/10.5281/zenodo.1467128}.

\bibitem[Wang(2009)]{wang09}
X.~Wang, (2009).
\newblock User manual for comcot version 1.7.
\newblock Technical report, Institute of Geological \& Nuclear Science, New
  Zealand.

\bibitem[Watkinson and Hall(2017)]{watkinson17}
I.~M. Watkinson and R.~Hall, (2017).
\newblock Fault systems of the eastern {Indonesian triple junction: Evaluation}
  of quaternary activity and implications for seismic hazards.
\newblock \emph{Geological Society, London, Special Publications},
  441:\penalty0 71--120.
\newblock
  \href{https://doi.org/10.1144/SP441.8}{https://doi.org/10.1144/SP441.8}.

\end{thebibliography}

\printindex


\end{document}